\documentclass[twocolumn,aps,prd,a4paper,final,superscriptaddress,longbibliography,showpacs,showkeys,nofootinbib]{revtex4-2}

\usepackage{amsmath,amssymb}
\usepackage{ascmac}
\usepackage{mathrsfs}
\usepackage[T1]{fontenc}
\usepackage{physics}
\usepackage{bm}
\usepackage{braket}
\usepackage{slashed}
\usepackage{float}
\usepackage{natbib}
\usepackage{mathtools}
\usepackage{comment}

\usepackage{graphicx}
\usepackage{hyperref}

\begin{document}

\newcommand{\ctext}[1]{\raise0.2ex\hbox{\textcircled{\scriptsize{#1}}}}
\renewcommand{\figurename}{Fig. }
\renewcommand{\thefigure}{\arabic{figure}}
\renewcommand{\tablename}{Table.}
\renewcommand{\thetable}{\arabic{table}}

\newcommand{\CP}{\mathbb{C}\mathrm{P}}

\twocolumngrid

%%%%%%%%%%%%%%%%%%%%%%%%%%%%%%%%%%%%%%%%%%%%%%%%%%%%%%%%%%%%%%%%%%%%%%
% TITLE

\title{Spectral flow of fermions in the $\CP^2$ (anti-)instanton, 
and the sphaleron with vanishing topological charge}

%%%%%%%%%%%%%%%%%%%%%%%%%%%%%%%%%%%%%%%%%%%%%%%%%%%%%%%%%%%%%%%%%%%%%%
%AUTHORS

\author{Yuki Amari}
\email{amari.yuki@keio.jp}
\affiliation{Research and Education Center for Natural Sciences, Keio University, Hiyoshi 4-1-1, Yokohama, Kanagawa 223-8521, Japan}
\affiliation{Department of Physics, Keio University, Hiyoshi 4-1-1, Yokohama, Kanagawa 223-8521, Japan}

\author{Nobuyuki Sawado}
\email{sawadoph@rs.tus.ac.jp}
\affiliation{Department of Physics and Astronomy, Faculty of Science and Technology, Tokyo University of Science, Noda, Chiba 278-8510, Japan}

\author{Shintaro Yamamoto}
\email{shintaroyamamoto019@gmail.com}
\affiliation{Department of Physics and Astronomy, Faculty of Science and Technology, Tokyo University of Science, Noda, Chiba 278-8510, Japan}

\vspace{.5 in}
\small

\date{\today}

%%%%%%%%%%%%%%%%%%%%%%%%%%%%%%%%%%%%%%%%%%%%%%%%%%%%%%%%%%%%%%%%%%%%%
% ABSTRACT

\begin{abstract}
The spectral flow is ubiquitous in the physics of soliton-fermion interacting systems. 
We study the spectral flows related to a continuous deformation of background soliton solutions, 
which enable us to develop insight into the emergence of fermionic zero modes and the localization mechanism of fermion densities.
We investigate a $\CP^2$ nonlinear sigma model  
in which there are the (anti-) instantons and also the sphalerons with vanishing topological charge. 
The standard Yukawa coupling of the fermion successfully generates infinite towers of the spectra and 
the spectral flow is observed when increasing the size of such solitons. At that moment, 
the localization of the fermions on the solitons emerges. The avoided crossings are also observed 
in several stages of the exchange of the flows, they are indicating a manifestation of the fermion exchange of the localizing nature.

\end{abstract}
%%%%%%%%%%%%%%%%%%%%%%%%%%%%%%%%%%%%%%%%%%%%%%%%%%%%%%%%%%%%%%%%%%%%%%

\maketitle 

%%%%%%%%%%%%%%%%%%%%%%%%%%%%%%%%%%%%%%%%%%%%%%%%%%%%%%%%%%%%%%%%%%%%%%
%INTRO

\section{Introduction}
\label{Sec:Intro}

Fermion-soliton interacting systems have been extensively investigated in a vast range of physics, 
including condensed matter physics, particle physics, planetary science and also cosmology. 
The normalizable modes with zero eigenenergy, the zero modes, certainly develop when fermions interact with soliton solutions
~\cite{Jackiw:1975fn,Jackiw:1981ee}
\footnote{It depends on the coupling scheme between fermions and solitons. 
In the fermion-vortex system, for example, the zero mode does not appear in the gauge coupling \cite{deVega:1976rt} 
while it does in the Yukawa coupling~\cite{Jackiw:1981ee}.}, 
as indicated by the Atiyah-Singer index theorem~\cite{Atiyah:1963zz,atiyah_patodi_singer_1975,PhysRevD.30.809,ANGHEL19891,cmp/1103904396}. 

The study of the topological solitons has been activated in the context of the fermion-soliton systems. 
In the $\phi^4$ kink, localized zero energy fermions and their fractionalization occur
~\cite{Dashen:1974cj,Jackiw:1975fn,PhysRevD.100.105003,PhysRevD.101.021701}. 
There are the normalizable zero modes in the vortex-fermion system~\cite{Jackiw:1981ee,Weinberg:1981eu}, 
and the 't Hooft-Polyakov magnetic monopoles
~\cite{Dashen:1974ck,Jackiw:1975fn,Callan:1982ac,Rubakov:1982fp,Rubakov:1988aq,Brennan:2021ewu,Brennan:2021ucy}. 
Another non-trivial effect is the nonconservation of the fermion numbers. 
The instanton in the Yang-Mills theory on four-dimensional Euclidean space~\cite{BELAVIN197585} 
has the nontrivial vacuum structure and gives rise to an anomalous phenomenon 
i.e., the fractionalization of the fermion number and the electric charge~\cite{Jackiw:1977pu}. 
The sphalerons~\cite{Klinkhamer:1984di,Manton:2004tk} in the Weinberg--Salam theory play a similar role to the instantons, 
leading to the nontrivial consequence~\cite{Kunz:1994ah}. 

In the case of the broken time reversal invariance, positive/negative energy eigenvalues 
of the fermions is asymmetric, which characterizes the spectral asymmetry~$N_\textrm{as}$
~\cite{Niemi:1984dx,Niemi:1984vz,PhysRevD.30.809,PhysRevLett.51.2077} 
\begin{align}
N_\textrm{as}:=\sum_{\varepsilon_{\mu}<0}1-\sum_{\varepsilon_{\mu}>0}1~\,,
\nonumber 
\end{align} 
where $\varepsilon_\mu$ are a tower of energy eigenvalues of fermions.
The fermion number $N_f$ in the fermion-soliton system is related to the spectral asymmetry as $N_f:=\frac{1}{2}N_\textrm{as}$.
The index theorem tells that the fermion number is equal to the topological charge $Q$ up to the sign, i.e.,
$N_f\propto Q$~\cite{PhysRevD.30.809,PhysRevLett.51.2077}. 
The spectral flow of the fermion energy eigenvalues originates from the coincidence.

Numerous details about the number of fermionic zero modes are provided by the index theorem, 
but it offers less information about the spectral flow, which is a transition in the spectrum 
with a changing parameter of the soliton in the background. 
In the instanton or the sphaleron background, rearrangements of energy spectra occur, e.g. one negative energy level 
undergoes the transition to a positive energy level concerning the imaginary time evolution~\cite{Kunz:1994ah}. 
Such behavior is commonly referred to as the spectral flow~\cite{Witten:2015aba,MANTON1985220,Alkofer:1995mv},
which is a manifestation of the anomaly~\cite{Witten:1982fp,Witten:2015aba,Niemi:1984di}. 
In addition to gauge theories, nonlinear sigma models have also been studied
~\cite{Amari:2019tgs,Amari:2023gjq,Delsate:2011aa,Kodama:2008xm,Perapechka:2018yux}.
The spectral flow argument is a direct tool for elucidating the connection between topology and 
flow of the fermions~\cite{Witten:1982fp,Burnier:2006za,Weinberg:2012pjx,Niemi:1984di,Niemi:1984vz}. 

The bulk-edge correspondences in 2D topological insulators and 3D Weyl semimetals have been extensively studied 
in a vast number of literature~\cite{Hasan:2010xy,YU2017550,Teo:2010zb,Fu:2007moz}. The spectral flow argument 
arises at some stages in the research. 
A more astounding application can be found in the shallow water dynamics of the ocean sea~\cite{Delplace2017,Tauber:2023}. 

In this paper, we examine the spectral flow analysis with the $\CP^2$ nonlinear sigma model. 
The model possesses two types of analytical solutions: the (anti-)instanton solutions
~\cite{DAdda:1978vbw}, and also the shpalerons, which are non-interacting mixtures of
instanton and anti-instanton~\cite{Din:1980jg,Din:1980uj,Din:1980wh}. 
There are several possibilities to incorporate the fermionic degrees of freedom into the model, e.g., 
the minimal coupling~\cite{Din:1982dh} and the supersymmetric coupling~\cite{DIN1981166}. 
In the present paper, we investigate a Yukawa coupling of the Dirac fermions with the instanton 
and also the sphaleron solutions. 
These solutions allow us to look at two significant issues. 
As we mentioned, the normalizable zero-mode solutions are caused by the finite charged topological solitons. 
Now, we shall study the spectral flow of Dirac fermions interacting with the net-zero topological charged solutions. 
Another topic is that we shall study \textit{the full} spectral flow analysis using the moduli parameters of the solutions.
In Refs.\cite{Kahana:1984be,Kahana:1984dx}, a size parameter was introduced for realizing the flow of the fermions, 
which has nothing to do with the solution itself. 
Another way is that one can employ an interpolating field 
between the soliton and the vacuum configuration 
$\phi_{\mathrm{int}}=\qty(1-\lambda)\phi_{\mathrm{vac}}+\lambda \phi_{\mathrm{sol}},\ 0\leq\lambda\leq1$, 
where the $\phi_{\mathrm{sol}}$ and $\phi_{\mathrm{vac}}$ denotes the soliton and the vacuum configuration
~\cite{Witten:1982fp,Amari:2019tgs}. 
One successfully obtains the normalizable modes, but $\phi_{\mathrm{int}}$ 
is only the exact solution at $\lambda=1$. 
In terms of the whole range of the moduli parameters of the solutions, 
we obtain the spectral flow with the true background soliton solutions. As a result, 
we investigate the sequences of the flows as growing the parameters. 
At the moment, we ignore any back-reactions from the fermions to the soliton solutions, although
the effects could be crucial for the structure of the solitons and also the spectral flows
~\cite{Perapechka:2018yux,Perapechka:2019upv,Loginov:2023rda,Loginov:2024nmi}. 

This paper is organized as follows: In Sec.~\ref{Sec:NLSM}, we introduce the %$\CP^{N-1}$ nonlinear sigma model and the 
solutions of the (anti-)instantons and the sphalerons. The topological charge is defined in this section.  
Their moduli parameters are described in Sec.~\ref{SubSec:Parameter}.  
Sec.~\ref{Sec:FSM} is the fermionic model and the details of our numerical method. 
In Sec.~\ref{Sec:Result_flow}, we present the results of the spectral flows for the anti-instanton and the mixture.  
In Sec.~\ref{Sec:Result_Density}, we are concentrating on the localization of the fermion density through the avoided crossing. 
Sec.~\ref{Sec:Conclusions} is devoted to our conclusions.

%%%%%%%%%%%%%%%%%%%%%%%%%%%%%%%%%%%%%%%%%%%%%%%%%%%%%%%%%%%%%%%%%%%%%%

%%%%%%%%%%%%%%%%%%%%%%%%%%%%%%%%%%%%%%%%%%%%%%%%%%%%%%%%%%%%%%%%%%%%%%
%DIN-ZAKRZEWSKI INSTANTONS

\section{$\CP^{2}$ Nonlinear Sigma Model}
\label{Sec:NLSM}
A nonlinear sigma model (NLSM) with a target space which is the complex projective space $\CP^{N-1}$ 
is an important class of quantum field theories~\cite{DAdda:1978vbw}. 
In any $N\geq2$, the $\CP^{N-1}$ NLSM has 2D mimic of (anti-) instanton solutions in Yang--Mills gauge theory. 
Moreover, there are further solutions in $N>2$ that are a non-interacting mixture of instantons 
and anti-instantons~\cite{Din:1980jg,Din:1980uj,Din:1980wh}, that we simply refer to it as the mixture. 

Instantons in $d$-dimensional Euclidean space can be regarded as static topological solitons in $(d+1)$ dimensions. 
A model consisting of fermions and instantons can have two different prescriptions. 
One is that, as utilized in Refs.\cite{Kiskis:1978tb,Burnier:2006za,Niemi:1984di,Niemi:1984vz}, 
fermions couple with instantons in a region of space where the instantons localize. 
Another involves embedding solutions in $(d+1)$ dimensional spacetime. 
In this paper, we follow the latter. 
In the following, we concentrate on the case of the $N=3$, but the extensions of $N\geq 3$ are straightforward.

\subsection{The model and the formal solutions}
\label{SubSec:DZsol}

In this subsection, we provide a formal introduction of the model and its solution. 
The static energy of the %$\CP^{N-1}$ 
$\CP^2$ NLSM in 
$(2+1)$-dimensional space-time is given by
\begin{align}
    E=\int\dd[2]{x}\qty(D_jZ)^\dagger\cdot\qty(D_jZ),\ j=1,2
\end{align}
where %$Z=\qty(Z_1,Z_2,\cdots,Z_N)^T$
$Z=\qty(Z_1,Z_2,Z_3)^T$ is a three dimensional complex unit vector $Z^\dagger\cdot Z=1$, 
and $D_j=\partial_j-Z^\dagger\cdot\partial_j Z$ defines a covariant derivative. 
Since soliton solutions should be of finite energy, 
the field $Z$ at the spatial infinity must be a constant vector up to $\mathrm{U\qty(1)}$ phase. 
Due to the boundary condition and the identification of $\CP^2$: $Z\sim Z'=e^{i\alpha\qty(x)}Z$, 
$\mathbb{R}^2$ is compactified to $2$-sphere $S^2$,  
and then
$\pi_2\qty(\CP^2)=\mathbb{Z}$. The solutions of %$\CP^{N-1}$ 
$\CP^2$ NLSM are classified by an integer, so-called topological charge $Q$, which is defined as
\begin{align}
    Q=-\frac{i}{2\pi}\int\dd[2]{x}\epsilon_{jk}\qty(D_jZ)^\dagger\cdot\qty(D_kZ),\ j,k=1,2\label{Eq:Q_original}
\end{align}
where $\epsilon_{jk}$ is an antisymmetric tensor: $\epsilon_{12}=-\epsilon_{21}=1, \epsilon_{11}=\epsilon_{22}=0$.
In terms of the complex coordinates $x_\pm=x_1\pm ix_2$ for convenience, 
the energy and the topological charge can be rewritten as 
\begin{align}
    E
	&=\int\dd[2]{x}2\qty[\abs{D_+Z}^2+\abs{D_-Z}^2],\label{Eq:NLSM}
\end{align}
\begin{align}
    Q&=\frac{1}{2\pi}\int\dd[2]{x}2\qty[\abs{D_+Z}^2-\abs{D_-Z}^2]\label{Eq:Q}
\end{align}
with $D_\pm=\partial_\pm-Z^\dagger\cdot\partial_\pm Z$.
The equation of motion (EOM) is 
\begin{align}
    D_-D_+Z+\abs{D_+Z}^2Z=0 
    . \label{Eq:EOM}
\end{align}
From Eq.~\eqref{Eq:NLSM} and Eq.~\eqref{Eq:Q}, we get
\begin{align}
    E
    &=2\pi \abs{Q}+4\int\dd[2]{x}\abs{D_\mp Z}^2\geq2\pi\abs{Q}\,.\label{Eq:Inequality}
\end{align}
The equality is realized when the solution satisfies the Bogomol'nyi-Prasad-Sommerfield (BPS) equation
\begin{equation}
    D_\mp Z=0.\label{Eq:BPS}
\end{equation}

%%%%%%%%%%%%%%%%%%%%%%%%%%%%Fig.1%%%%%%%%%%%%%%%%%%%%%%%%%%%%

\begin{figure*}[t]
	\hspace*{-0.06\linewidth}
	\centering
	\includegraphics[width=1.08\linewidth]{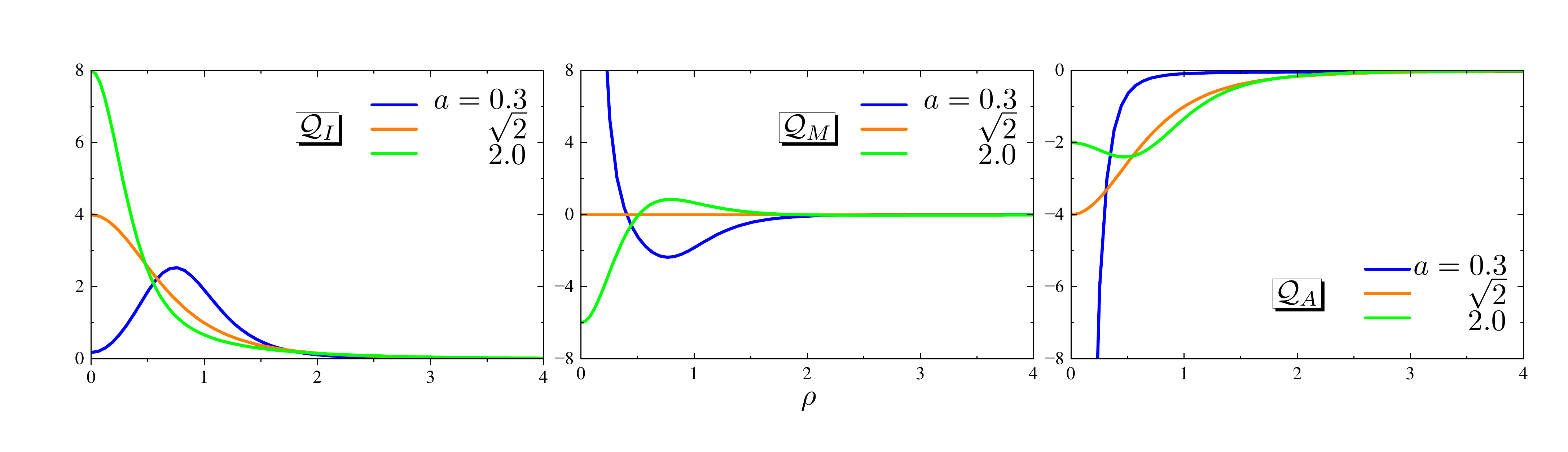}
	\caption{The $a$ dependences of $\mathcal{Q}_I$, $\mathcal{Q}_A$ and $\mathcal{Q}_M$ with $b=1$. 
	Although $\mathcal{Q}_A$ and $\mathcal{Q}_M$ have a blue curve ($a=0.3$) that appears to be divergent, 
	they are actually finite at the origin, i.e, $\mathcal{Q}_A\qty(0)\sim -88$ and $\mathcal{Q}_M\qty(0)\sim 88$, respectively.}
	\label{Fig:Qb1}
\end{figure*}
\if0{
\begin{figure}[t]
	\begin{minipage}[b]{0.32\linewidth}
		\centering
  \hspace*{-6ex}
		\includegraphics[keepaspectratio, scale=0.24]{QIb1v3.pdf}
	\end{minipage}
	\begin{minipage}[b]{0.32\linewidth}
		\centering
  \hspace*{-3ex}
		\includegraphics[keepaspectratio, scale=0.24]{QMb1v3.pdf}
	\end{minipage}
	\begin{minipage}[b]{0.32\linewidth}
		\centering
		\includegraphics[keepaspectratio, scale=0.24]{QAb1v3.pdf}
	\end{minipage}
	\caption{The $a$ dependences of $\mathcal{Q}_I$, $\mathcal{Q}_A$ and $\mathcal{Q}_M$ with $b=1$. The blue curve ($a=0.3$) of $\mathcal{Q}_A$ and $\mathcal{Q}_M$ may look divergent, but their maximum value are finite, i.e, $\mathcal{Q}_A\qty(0)=-88$ and $\mathcal{Q}_M\qty(0)=88$ respectively.}
	\label{Fig:Qb1}
\end{figure}
}\fi

%%%%%%%%%%%%%%%%%%%%%%%%%%%%Fig.1%%%%%%%%%%%%%%%%%%%%%%%%%%%%

The solutions of the equation are referred to as the (anti-)instantons.
%The $\CP^2$ model possesses instanton solutions. 
Their forms can be written in terms of holomorphic vectors
\begin{align}
	Z^{\pm}=\frac{f\qty(x_\pm)}{\abs{f\qty(x_\pm)}}
\end{align}
where $f\qty(x_\pm)$ is $3$-dimensional non-zero holomorphic vectors ($+:$ instanton, $-:$ anti-instanton).
The solutions automatically satisfy the BPS equation~(\ref{Eq:BPS}) and also the EOM, and 
can thus minimize the energy, i.e., $E=2\pi\abs{Q}$. Eqs.~\eqref{Eq:Q},~\eqref{Eq:BPS} tell us 
that the (anti-)instantons possess a positive (negative) %sign 
topological charge.
Furthermore, a more general class of solutions has been discovered by Din and Zakrzewski, 
that we refer to it as Din--Zakrzewski (DZ) solutions~\cite{Din:1980jg,Din:1980uj,Din:1980wh}. 
Using the B\"acklund transformation
\begin{align}
	P_+g=\partial_+ g-\frac{g^\dagger\cdot\partial_+ g}{\abs{g}^2}g,
\end{align}
with $g\in\mathbb{C}^3\backslash\qty{0}$, %$g\in\mathbb{C}^N\backslash\qty{0}$ is any non-zero vector, 
the DZ solutions are obtained as
\begin{align}
	Z_{\qty(k)}=\frac{P_+^kf}{\abs{P_+^kf}}\ \qty(k=0,\ldots,2).
\end{align}
Note that $Z_{\qty(k)}$s are independent solutions of the EOM~\eqref{Eq:EOM} for each $k$. One can verify that only $Z_{\qty(0)}$ and %$Z_{\qty(N-1)}$ 
$Z_{\qty(2)}$ satisfy the BPS equation, i.e., $D_-Z_{\qty(0)}=0$ and $D_+Z_{\qty(2)}=0$. %$D_+Z_{\qty(N-1)}=0$. 
This means that $Z_{\qty(0)}$ has positive, and %$Z_{\qty(N-1)}$ 
$Z_{\qty(2)}$ has negative topological charge. Therefore, they are \textit{instanton} and \textit{anti-instanton} solutions respectively. Generally, the other solution, $k=1$, does not satisfy the BPS equation. %Generally, the other solutions, $0<k<N-1$, do not satisfy BPS equation. 
From Eq.~\eqref{Eq:Inequality}, we can see that if $Z$ satisfies the BPS equation~\eqref{Eq:BPS}, the energy is minimized, whereas if it does not, the energy is higher. Therefore, such a non-BPS solution is a saddle point for the energy. As we see later, it can be interpreted as a mixture solution of instantons and anti-instantons.%Therefore, such non-BPS solutions are saddle points for the energy. As we see later, they can be interpreted as a mixture of solutions of instantons and anti-instantons.

Since there are three types of the solutions $\qty(Z_{\qty(0)},Z_{\qty(1)},Z_{\qty(2)})$ in the $\CP^2$ NLSM. 
For all solutions, we give formal forms of the energy, the topological charge, and their densities
\begin{align}
      E_{\qty(k)}
	&=\int\dd[2]{x}\mathcal{E}_{\qty(k)},\nonumber \\
      \mathcal{E}_{\qty(k)}&\coloneqq2\qty[\abs{D_+Z_{\qty(k)}}^2+\abs{D_-Z_{\qty(k)}}^2];\\
    	Q_{\qty(k)}
	&=\frac{1}{2\pi}\int\dd[2]{x}\mathcal{Q}_{\qty(k)},\nonumber \\
	\mathcal{Q}_{\qty(k)}&\coloneqq2\qty[\abs{D_+Z_{\qty(k)}}^2-\abs{D_-Z_{\qty(k)}}^2],\ k=0,1,2.
\end{align}
The $Z_{\qty(0)}$ and $Z_{\qty(2)}$ are the solutions of the BPS equation, but the $Z_{\qty(1)}$ is not. 
More explicit form of the energy densities and the topological charge densities are given as
\begin{align}
	\mathcal{E}_{\qty(0)}
        &=2\abs{D_+Z_{\qty(0)}}^2,~~
	\mathcal{E}_{\qty(2)}=2\abs{D_-Z_{\qty(2)}}^2,\nonumber \\
	\mathcal{E}_{\qty(1)}
      &=2\qty[\abs{D_+Z_{\qty(1)}}^2+\abs{D_-Z_{\qty(1)}}^2];\\
	        \mathcal{Q}_{\qty(0)}
        &=2\abs{D_+Z_{\qty(0)}}^2,
	~~\mathcal{Q}_{\qty(2)}=-2\abs{D_-Z_{\qty(2)}}^2,\nonumber \\
        \mathcal{Q}_{\qty(1)}
        &=2\qty[\abs{D_+Z_{\qty(1)}}^2-\abs{D_-Z_{\qty(1)}}^2].
 \end{align}
From these, one directly checks that the energy and the topological charge of the instanton and the anti-instanton
have the connection $E_{\qty(0)}=2\pi Q_{\qty(0)}$ and $E_{\qty(2)}=2\pi\abs{ Q_{\qty(2)}}$. 
Also, after some algebra 
(see the Appendix~\ref{AppendixA}), 
we can confirm that the components of $\mathcal{Q}_{\qty(1)}$ and $\mathcal{E}_{\qty(1)}$ can be written as 
\begin{align}
   \abs{D_+Z_{\qty(1)}}^2=\abs{D_-Z_{\qty(2)}}^2,\ \abs{D_-Z_{\qty(1)}}^2]=\abs{D_+Z_{\qty(0)}}^2.
\end{align}
As a result, the energy $E_{\qty(1)}$ and the topological charge $Q_{\qty(1)}$ have the form below:
\begin{align}
    E_{\qty(1)}&=E_{\qty(0)}+E_{\qty(2)};\label{Eq:ETotal}\\
    Q_{\qty(1)}&=\abs{Q_{\qty(2)}}+\qty(-Q_{\qty(0)})\label{Eq:QTotal}.
\end{align}
From these, $Z_{\qty(1)}$ can be interpreted as a non-interacting composite 
of the $Q_{\qty(0)}$ instantons and the $Q_{\qty(2)}$ anti-instantons. 
Therefore, we call the solution $Z_{\qty(1)}$ as the mixture. 
Since $\abs{Q_{\qty(2)}}$ and $\qty(-Q_{\qty(0)})$ are the positive and negative part of 
the topological charge of the mixture, then 
the anti-instanton $Z_{\qty(2)}$ and the instanton $Z_{\qty(0)}$ are the instanton and anti-instanton part of 
the mixture, respectively. 
To make the meaning more clear, we rewrite $\qty(Z_{\qty(0)},Z_{\qty(1)},Z_{\qty(2)})$ to $\qty(Z_I,Z_M,Z_A)$ 
which stand for the instanton, the mixture, and the anti-instanton solutions, respectively.
Accordingly, the energy, the topological charges and the densities are now $\qty(E_I,E_M,E_A)$, 
$\qty(Q_I,Q_M,Q_A)$, $\qty(\mathcal{E}_I,\mathcal{E}_M,\mathcal{E}_A)$ and $\qty(\mathcal{Q}_I,\mathcal{Q}_M,\mathcal{Q}_A)$.

\subsection{The explicit solutions}
\label{SubSec:Parameter}

In this subsection, we construct the explicit DZ solutions in 
the $\CP^2$ NLSM, with introducing the parameters which govern the deformation of solutions. 
We employ a simple holomorphic vector $f\qty(x_+)\in\mathbb{C}^3\backslash\qty{0}$
\begin{align}
    f=\mqty
    (
        1\\
        a\rho \exp\qty[i\phi]\\
        b\rho^2 \exp\qty[2i\phi]
    )
\end{align}
where $\qty(\rho,\phi)$ denote polar coordinates defined through $x_\pm=x_1\pm ix_2=\rho\exp\qty[\pm i\phi]$ and $\qty(a,b)$ are generally complex constants. 
Here we note that thanks to $\mathrm{U}\qty(3)$ global symmetry of the model~\eqref{Eq:NLSM}, they are set to be non-negative real numbers without loss of generality. 
For the boundary conditions imposed on the soliton solutions, the finiteness of $\qty(a,b)$ is required.

%%%%%%%%%%%%%%%%%%%%%%%%%%%%Fig.2%%%%%%%%%%%%%%%%%%%%%%%%%%%%

\begin{figure*}[t]
	\hspace*{-0.06\linewidth}
	\centering
	\includegraphics[width=1.08\linewidth]{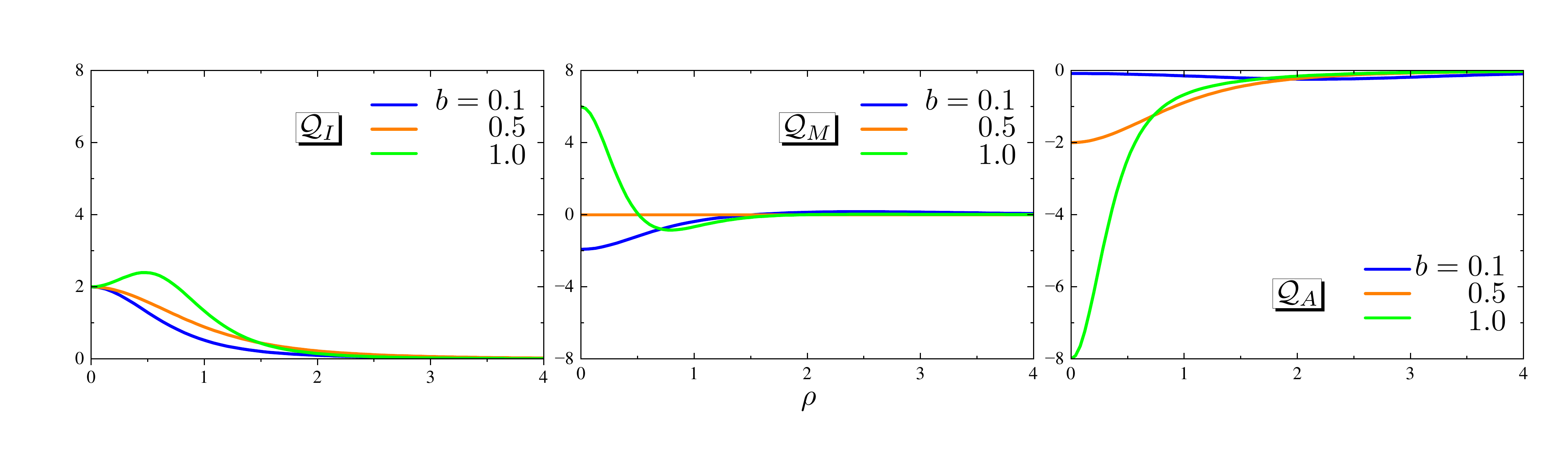}
	\caption{The $b$ dependences of $\mathcal{Q}_I$, $\mathcal{Q}_A$ and $\mathcal{Q}_M$ with $a=1$.}
	\label{Fig:Qa1}
\end{figure*}

%%%%%%%%%%%%%%%%%%%%%%%%%%%%Fig.2%%%%%%%%%%%%%%%%%%%%%%%%%%%%

By the nonlinear operator $P_+$ acting on $f$, we obtain the DZ solutions $\qty(Z_I,Z_M,Z_A)$ as followings:
\begin{align}
        Z_I&=\frac{f}{\abs{f}}=\frac{1}{\sqrt{\Delta_1}}\mqty
	(
		1\\
		a\rho\exp\qty[i\phi]\\
		b\rho^2\exp\qty[2i\phi]
	),\label{Eq:ZI}\\
	Z_M&=\frac{P_+f}{\abs{P_+f}}=\frac{1}{\sqrt{\Delta_1\Delta_2}}\mqty
	(
		-\qty(a^2+2b^2\rho^2)\rho\exp\qty[-i\phi]\\
		a\qty(1-b^2\rho^4)\\
		b\qty(2+a^2\rho^2)\rho\exp\qty[i\phi]
	),\label{Eq:ZM}\\
	Z_A&=\frac{P_+^2f}{\abs{P_+^2f}}=\frac{1}{\sqrt{\Delta_2}}\mqty
	(
		ab\rho^2\exp\qty[-2i\phi]\\
		-2b\rho\exp\qty[-i\phi]\\
		a
	).\label{Eq:ZA}
\end{align}
where 
$\qty(\Delta_1,\Delta_2)$ are defined by
\begin{align}
        \Delta_1&\equiv\abs{f}^2=1+a^2\rho^2+b^2\rho^4,\nonumber\\
        \Delta_2&\equiv\abs{f}^2\abs{P_+f}^2=a^2+4b^2\rho^2+a^2b^2\rho^4.\nonumber
\end{align}
Here we emphasize that, for arbitrary parameters $\qty(a,b)\in\qty(0,\infty)$,
$Z_I$, $Z_M$ and $Z_A$ are exact solutions of EOM~\eqref{Eq:EOM}.
We designate Eq.\eqref{Eq:ZI}--Eq.\eqref{Eq:ZA} as axially symmetric solutions because 
the topological charge densities and energy densities evaluated 
by \eqref{Eq:ZI}--\eqref{Eq:ZA} are the function of the radial 
coordinate $\rho$. 

\renewcommand{\arraystretch}{2.5}
\begin{table}[H]
	\centering
    \caption{Values of the topological charges at the boundaries of the parameter limits.  
}
    \label{Table:Qs}
		\begin{tabular}{l|ccc}
			\hline\hline
            The parameters&~~$Q_{I}$~~&~~$Q_{M}$~~&~~$Q_{A}$~~\\
			\hline
			$a=0,~~b=\mathrm{finite}~~$&$2$&$-2$&$0$\\
			$\dfrac{1}{a}=0,~~b=\mathrm{finite}$~~&$0$&$2$&$-2$\\
			$a=\mathrm{finite},~~b=0$~~&$1$&$-1$&$0$\\
			$a=\mathrm{finite},~~\dfrac{1}{b}=0$~~&$0$&$1$&$-1$\\
			\hline\hline
		\end{tabular}
\end{table}
\renewcommand{\arraystretch}{1.0}
The topological charge densities can be written as 
\begin{align}
	\mathcal{Q}_I&=2\frac{\Delta_2}{\qty(\Delta_1)^2},\label{Eq:QI}\\
	\mathcal{Q}_M&=2\qty(4a^2b^2\frac{\Delta_1}{\qty(\Delta_2)^2}-\frac{\Delta_2}{\qty(\Delta_1)^2}),\label{Eq:QM}\\
	\mathcal{Q}_A&=-2\cdot4a^2b^2\frac{\Delta_1}{\qty(\Delta_2)^2}\label{Eq:QA}.
\end{align}
After the integration, we obtain the topological charges
\begin{align}
    Q_I=2,\ Q_M=0,\ Q_A=-2.\label{Eq:Q_normal}
\end{align}
which is supported for all the parameters choice $\qty(a,b)\in\qty(0,\infty)$. 
It should be emphasized that we obtain completely different values for the topological charges 
when we use the limiting values of the parameters, such as $\dfrac{1}{a}=0$. 
The results are summarized in Table.~\ref{Table:Qs}. As a result,  
the deformations of the solutions successfully induce the spectral flow 
since the DZ solutions~\eqref{Eq:ZI}-\eqref{Eq:ZA} are the function of the parameters. 
For example, a parameter sequence $a:0\to \mathrm{finite}\to \infty$ 
with finite $b$ of the anti-instanton corresponds to a transition of the anti-instanton charge $Q_A:0\to-2\to-2$,   
that certainly causes the two levels to cross the zero energy line as $a$ grows.

%%%%%%%%%%%%%%%%%%%%%%%%%%%%%%%%%%%%%%%%%%%%%%%%%%%%%%%%%%%%%%%%%%%%%%

%%%%%%%%%%%%%%%%%%%%%%%%%%%%%%%%%%%%%%%%%%%%%%%%%%%%%%%%%%%%%%%%%%%%%%
%THE FERMIONIC MODEL

\section{Fermionic sigma model}
\label{Sec:FSM}

We consider the fermions coupled with the DZ solutions that are embedded into $(2+1)$ dimensional spacetime. 

\subsection{The Effective Action and the NLSM}
For the construction of the model, we begin with a partition function $\mathcal{Z}$ 
defined in three dimensional Euclidean space 
\begin{align}
        \mathcal{Z}
	&=\int\mathcal{D}Z\mathcal{D}\Psi\mathcal{D}\overline{\Psi}\exp\qty[-\int\dd[3]{x}\overline{\Psi}i\mathfrak{D}\Psi].
\end{align}
where $i\mathfrak{D}\equiv i\gamma_\mu\partial_\mu+mX$ is called a Dirac operator. 
Here $m\in\mathbb{R}$ is a coupling constant and the gamma matrices $\gamma_\mu$ are given by the standard Clifford algebra
\begin{align}
	\qty{\gamma_\mu,\gamma_\nu}&=-2\delta_{\mu\nu}I_2.
\end{align}
In the two space dimensions, 
the matrices are given by $\gamma_1=i\sigma_1,\ \gamma_2=i\sigma_2,$ and $\gamma_3=-i\sigma_3$ where $\sigma_j$ are the 
standard Pauli matrices.
For the spectral flow analysis of the DZ solutions, we provide an efficient Yukawa-type coupling 
with a nice parametrization of $\CP^2$ in terms of a matrix-valued field $X$, 
i.e., the principal variable.~\cite{Eichenherr:1979hz,Ferreira:2010jb}. 
Formally, it is defined by
\begin{align}
	X(g)=g\sigma(g)^{-1},~~~~g\in \textrm{SU}(3)\,.
	\label{pvariable}
\end{align}
It parametrizes the coset space $\CP^{2}=\mathrm{SU}(3)/\mathrm{SU}(2)\otimes \mathrm{U}(1)$ with the subgroup 
$\mathrm{SU}(2)\otimes \mathrm{U}(1)$ being invariant under the involutive automorphism $(\sigma^2=1)$.
It follows that $X(gk)=X(g)$ for $\sigma(k)=k,~~k\in \mathrm{SU}(2)\otimes \mathrm{U}(1)$. 
The $\CP^{2}$ model possesses some symmetries.  
With the definition~(\ref{pvariable}), one finds that the principal variable possesses the following property
\begin{align}
	X^\dagger=X^{-1}=X,\ X^2=I_3.
\end{align}

The low-energy effective action of this theory includes the aforementioned $\CP^2$ NLSM, which can be derived 
using a perturbative method. 
Here, we give a brief outline of the derivation using a normal derivative expansion~\cite{Diakonov:1987ty,Abanov:1999qz}. 
Integrating out the fermion fields in the partition function $\mathcal{Z}$, we obtain so-called a Dirac determinant
\begin{align}
    \int\mathcal{D}\Psi\mathcal{D}\overline{\Psi}\exp\qty[-\int\dd[3]{x}\overline{\Psi}i\mathfrak{D}\Psi]=\det\qty[i\mathfrak{D}].
\end{align}
Therefore, $\mathcal{Z}$ is now
\begin{align}
	\mathcal{Z}\propto\int\mathcal{D}Z\exp\qty[-S\qty[X]],~~S\qty[X]=-\ln\det\qty[i\mathfrak{D}],
\end{align}
where $S$ is called the effective action. 
$i\mathfrak{D}$ is not Hermitian in Euclidean space, so we separate $S$ into the real part and the imaginary part, 
$S=\Re S+i\Im S$ and each component becomes 
\begin{align}
        \Re S&=-\frac{1}{2}\mathrm{Sp}\ln \mathfrak{D}^\dagger \mathfrak{D},\\
	\Im S&=-\frac{1}{2i}\mathrm{Sp}\ln\frac{i\mathfrak{D}}{\qty(i\mathfrak{D})^\dagger}
\end{align}
where $\mathrm{Sp}$ means a functional trace taking all trace, i.e. matrices, states, and Euclidean space. 
Using a derivative expansion, we can evaluate the regularized real part of the action leaving only the first nonzero orders in $\frac{1}{m}$ as
\begin{align}
        \Re S
	&=\frac{m}{16\pi}\int\dd[3]{x}\mathrm{Tr}\qty(\partial_\mu X\partial_\mu X)+\order{m^0}.\label{Eq:EffModel}
\end{align}
Note that the principal variable has directly related to $Z$,
\footnote{Through this representation, 
the principal variable $X$ has another direct connection with the $\mathrm{SU}(3)$ 
color field $\mathfrak{n}$, such that $\mathfrak{n}=\frac{1}{3}I_3-X$.  
For more general case of $\CP^{N-1}$, this kind of connection holds.} 
such as
\begin{align}
	X= I_3-2Z\otimes Z^\dagger,\ I_3=\mqty(1&0&0\\0&1&0\\0&0&1). 
\end{align}
From this, we confirm that the action (\ref{Eq:EffModel}) becomes 
\begin{align}
	\Re S
	&=\frac{m}{2\pi}\int\dd{x_3}E\qty[Z]+\order{m^0}
\end{align}
where $E(Z)$ is exactly the energy functional in the $\CP^2$ NLSM. 
Since the background field $Z$ is independent of $x_3$, the terms with imaginary time derivatives vanish. 
We can evaluate the imaginary part of the action in a similar fashion, which often contains the 
imaginary time derivative, and always vanishes in our model. 

\subsection{The Dirac Hamiltonian and Method for the Computation}

For the spectral flow analysis, we numerically solve an eigenequation of the Dirac Hamiltonian $\mathcal{H}$
\begin{align}
	\mathcal{H}\psi\qty(\bm{x})=\varepsilon\psi\qty(\bm{x}).\label{Eq:Eigen}
\end{align}
The Hamiltonian is derived from the Dirac operator
\begin{align}
    i\mathfrak{D}=i\gamma_3\qty(\partial_3+\mathcal{H}).
\end{align}
The explicit form of the Hamiltonian is 
\begin{align}
	\mathcal{H}&=i\gamma_3\qty(i\gamma_k\partial_k+mX)\nonumber\\
	&=\mqty
	(
		mX&\exp\qty[-i\phi]\qty(-\partial_\rho+\frac{i\partial_\phi}{\rho})\\
		\exp\qty[i\phi]\qty(\partial_\rho+\frac{i\partial_\phi}{\rho})&-mX
	).
\label{hamiltonian}
\end{align}
The Hamiltonian is commuted with an extended spin operator 
$\mathcal{K}$, so-called grand-spin operator~\cite{Amari:2019tgs}, 
which is defined by 
\begin{align}
    \mathcal{K}=l_3+\frac{1}{2}\sigma_3+\frac{1}{2}\lambda_3+\frac{\sqrt{3}}{2}\lambda_8\label{Eq:K}
\end{align}
where $l_3$, $\lambda_3$ and $\lambda_8$ represent an angular momentum operator, 
the third and eighth components of Gell--Mann matrices respectively. 
This operator satisfies $\mathcal{K}\mathcal{H}\mathcal{K}^{-1}=\mathcal{H}$.
Therefore, the eigenvalue $\kappa$ is conserved and labels the energy levels and the fermion densities. 
Also, irrespective of details of the background fields~\eqref{Eq:ZI}--\eqref{Eq:ZA}, 
when the fields are axially symmetric, the operator ~\eqref{Eq:K} always commutes with the Dirac Hamiltonian.

%%%%%%%%%%%%%%%%%%%%%%%%%%%%Fig.3%%%%%%%%%%%%%%%%%%%%%%%%%%%%
\begin{figure*}[t]
	\centering
		\includegraphics[width=0.5\linewidth]{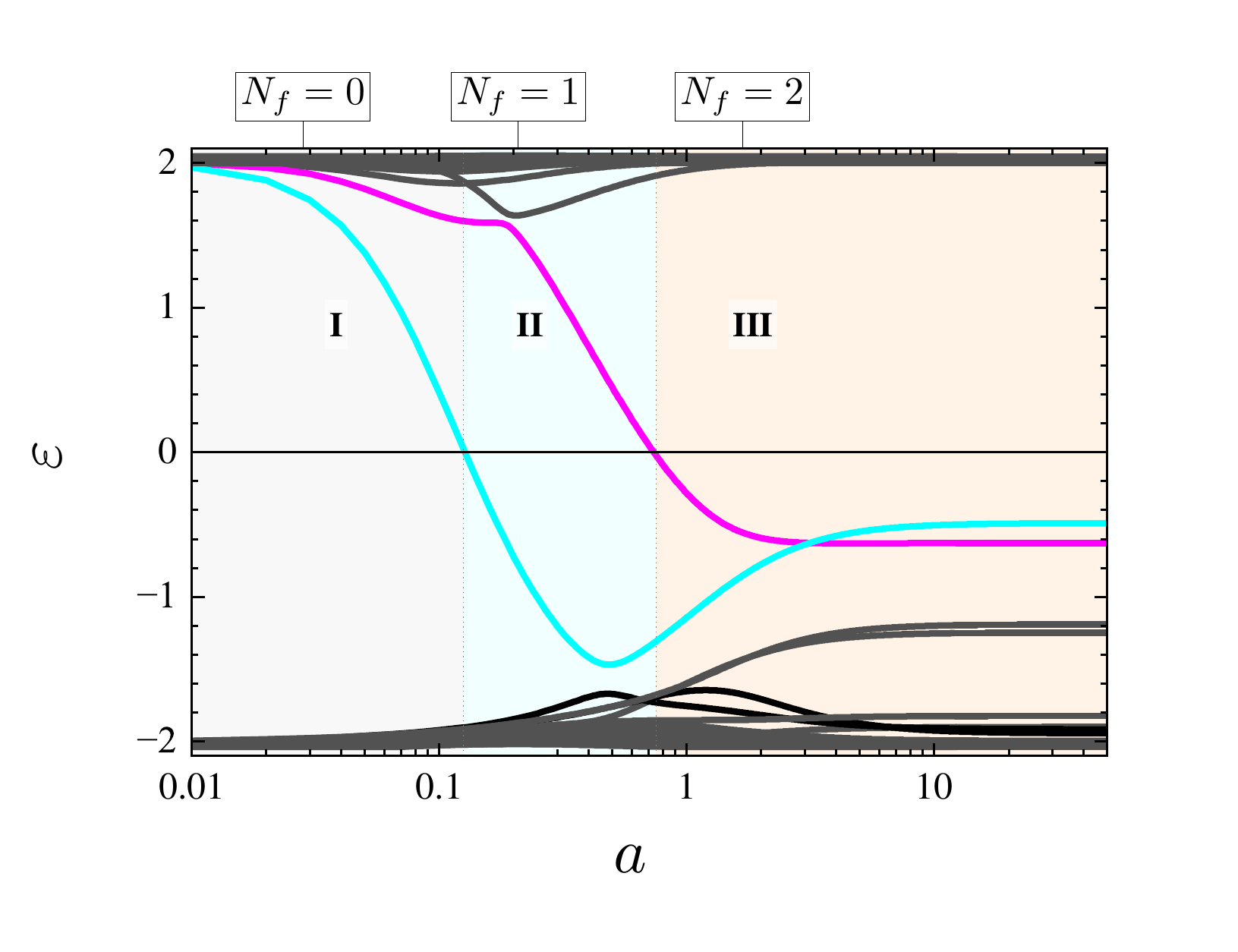}\hspace{-0.5cm}
		\includegraphics[width=0.5\linewidth]{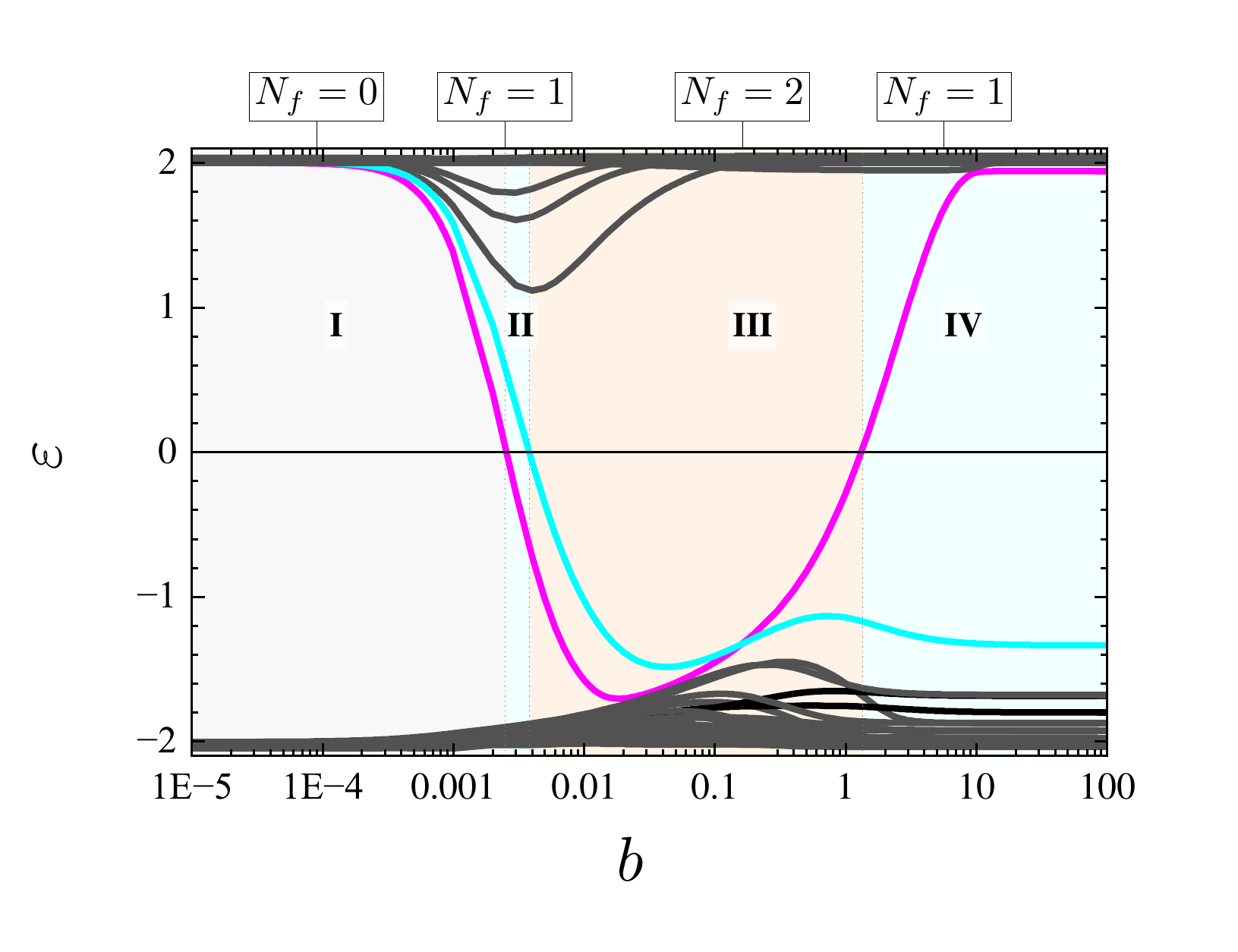}
		\\
		(a) \hspace{8cm}(b)
		\caption{\label{Fig:ZAflow}The spectral flow of the anti-instanton solution. 
		We illustrate the colored cyan and magenta lines as the crossing levels. 		
		(a)~The flow with change of the parameter $a$ with fixed $b=1.0$, and 
		(b)~of the parameter $b$ with $a=1.0$. }
\end{figure*}

%%%%%%%%%%%%%%%%%%%%%%%%%%%%Fig.3%%%%%%%%%%%%%%%%%%%%%%%%%%%%

%%%%%%%%%%%%%%%%%%%%%%%%%%%%Fig.4%%%%%%%%%%%%%%%%%%%%%%%%%%%%
\begin{figure*}[tbp]
	\centering
		\includegraphics[width=0.5\linewidth]{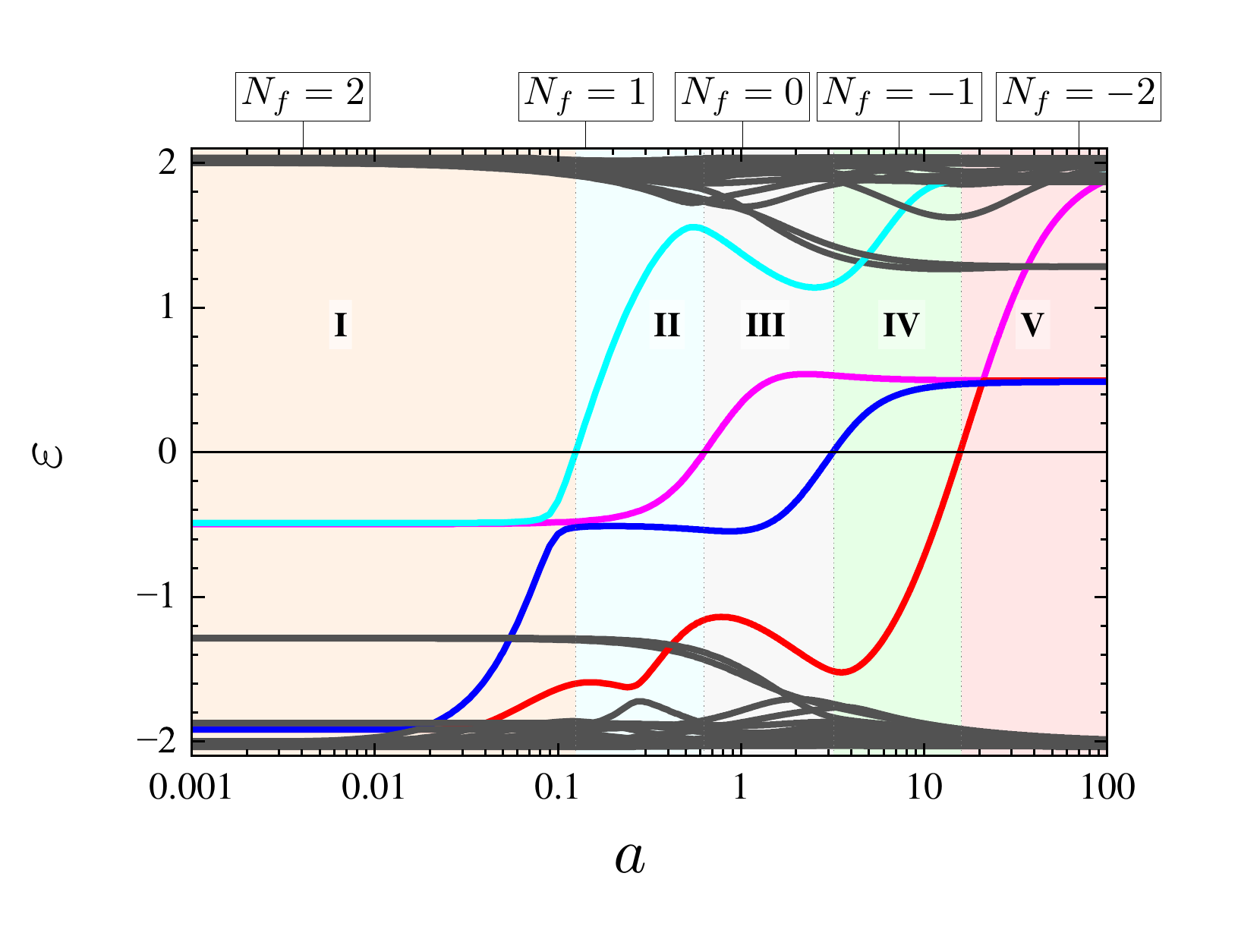}\hspace{-0.5cm}
		\includegraphics[width=0.5\linewidth]{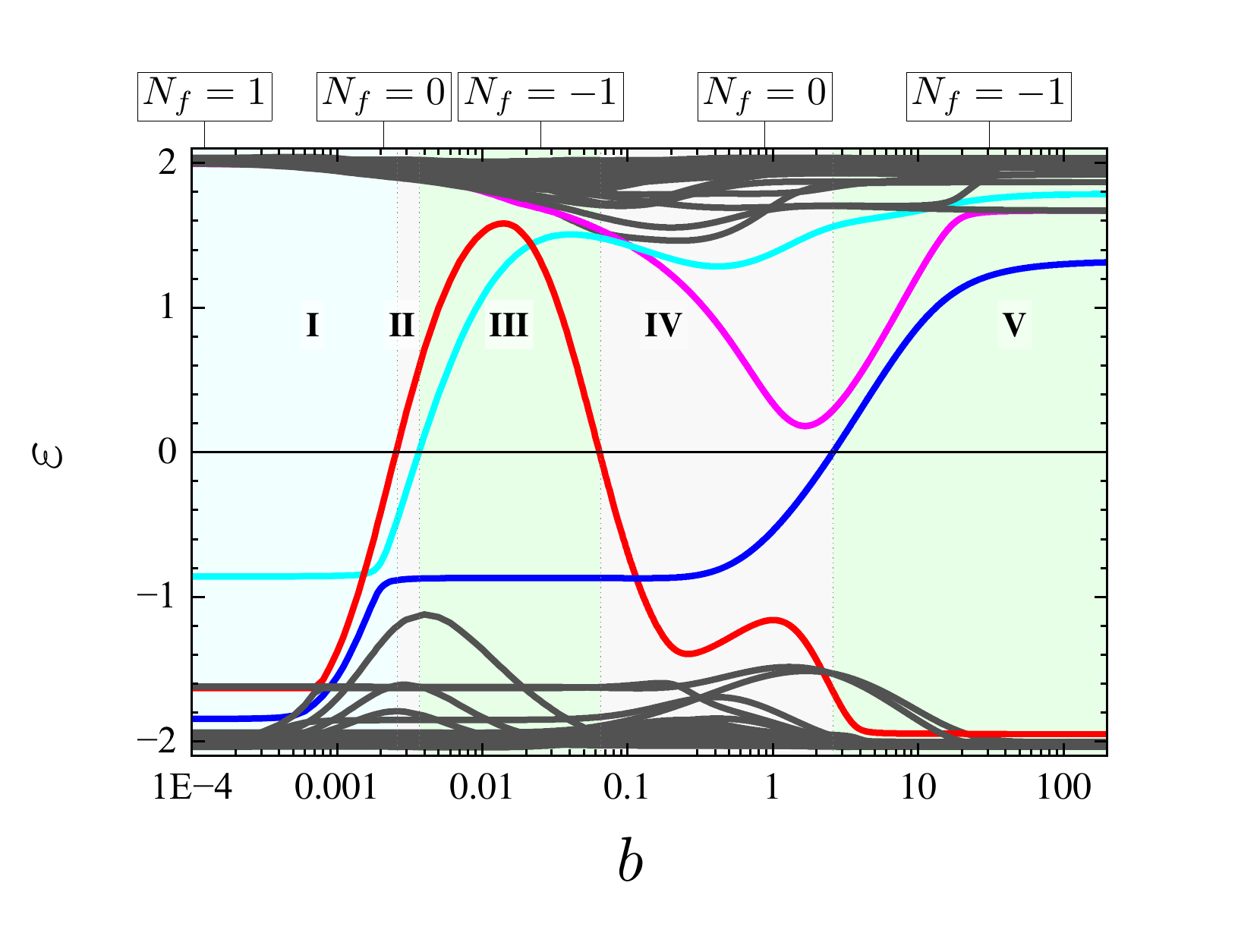}
		\\
		(a) \hspace{8cm}(b)
		\caption{\label{Fig:ZMflow}The spectral flow of the mixture solution. 
		We illustrate the colored red, blue, cyan and magenta lines as the 
		crossing levels. 
		(a)~The flow with change of the parameter $a$ with fixed $b=1.0$, and 
		(b)~of the parameter $b$ with $a=1.0$.}
\end{figure*}
%%%%%%%%%%%%%%%%%%%%%%%%%%%%Fig.4%%%%%%%%%%%%%%%%%%%%%%%%%%%%

%%%%%%%%%%%%%%%%%%%%%%%%%%%%%%%%%%%%%%%%%%%%%%%%%%%%%%%%%%%%%%%%%%%%%%

%%%%%%%%%%%%%%%%%%%%%%%%%%%%%%%%%%%%%%%%%%%%%%%%%%%%%%%%%%%%%%%%%%%%%%

We solve the Dirac eigenequation~\eqref{Eq:Eigen} corresponding to the Hamiltonian (\ref{hamiltonian}) numerically. 
Since the Hamiltonian depends on the parameters $\qty(a,b)$ through the $X$, 
we can analyze the spectral flow of the fermions as a function of $\qty(a,b)$.

Here we provide a quick overview of the procedure that is being developed in Ref.\cite{Amari:2019agx}. 
The secular equation for the energy eigenvalue $\varepsilon$ is given by
\begin{align}
	\det\qty(H-\varepsilon \Tilde{I})=0,\label{Eq:Secular}
\end{align}
where the matrix element of $H$ and $\Tilde{I}$ is given by
\begin{align}
	H_{k'^{\qty(p)},k^{\qty(q)}}&\equiv\int\dd[2]{x}\Phi_{\kappa}^{\qty(p)}\qty(k'^{\qty(p)},x)^\dagger\mathcal{H}\Phi_{\kappa}^{\qty(q)}\qty(k^{\qty(q)},x),\\
	\Tilde{I}_{k'^{\qty(p)},k^{\qty(q)}}&\equiv\int\dd[2]{x}\Phi_{\kappa}^{\qty(p)}\qty(k'^{\qty(p)},x)^\dagger\Phi_{\kappa}^{\qty(q)}\qty(k^{\qty(q)},x).
\end{align}
To obtain these matrix representations, we construct plane-wave spinor bases 
$\Phi_\kappa^{(p)}$ which are the eigenvectors of the Hamiltonian of the massive free fermions $\mathcal{H}^0=i\gamma_3\qty(i\gamma_k\partial_k\pm m)$.
They are given by
\begin{align}
	\Phi_\kappa^{(u)}&=\mathcal{N}_i^{(u)}
	\mqty
	(
		\zeta_{+}^{(u)}J_{\kappa-\frac{3}{2}}(k_i^{(u)}\rho)\exp\qty[i\qty(\kappa-\frac{3}{2})\phi]\\
		\zeta_{-}^{(u)}J_{\kappa-\frac{1}{2}}(k_i^{(u)}\rho)\exp\qty[i\qty(\kappa-\frac{1}{2})\phi]
	)
	\otimes
	\mqty
	(
		1\\
		0\\
		0
	),\\
	\Phi_\kappa^{(d)}&=\mathcal{N}_i^{(d)}
	\mqty
	(
		\zeta_{-}^{(d)}J_{\kappa-\frac{1}{2}}(k_i^{(d)}\rho)\exp\qty[i\qty(\kappa-\frac{1}{2})\phi]\\
		\zeta_{+}^{(d)}J_{\kappa+\frac{1}{2}}(k_i^{(d)}\rho)\exp\qty[i\qty(\kappa+\frac{1}{2})\phi]
	)
	\otimes
	\mqty
	(
		0\\
		1\\
		0
	),\\
	\Phi_\kappa^{(s)}&=\mathcal{N}_i^{(s)}
	\mqty
	(
		\zeta_{+}^{(s)}J_{\kappa+\frac{1}{2}}(k_i^{(s)}\rho)\exp\qty[i\qty(\kappa+\frac{1}{2})\phi]\\
		\zeta_{-}^{(s)}J_{\kappa+\frac{3}{2}}(k_i^{(s)}\rho)\exp\qty[i\qty(\kappa+\frac{3}{2})\phi]
	)
	\otimes
	\mqty
	(
		0\\
		0\\
		1
	)
\end{align}
where $J_{l^{\qty(p)}}\qty(k_i^{(p)}\rho)$ are standard Bessel functions with the
integer order $l^{\qty(p)}$. The coefficients $\zeta$ are given by 
\begin{align}
	\zeta^{(p)}_{+}&=
    \begin{cases}
    \ \ \ \ \ \ \ 1&\qty(\omega_i^{\qty(p)}>0)\\
    \frac{k_i^{\qty(p)}}{\abs{\omega_i^{\qty(p)}}+m}&\qty(\omega_i^{\qty(p)}<0)
    \end{cases}\nonumber\\
    \zeta^{(p)}_{-}&=
    \begin{cases}
    -\frac{k_i^{\qty(p)}}{\omega_i^{\qty(p)}+m}&\qty(\omega_i^{\qty(p)}>0)\\
    \ \ \ \ \ \ \ 1&\qty(\omega_i^{\qty(p)}<0)
    \end{cases}\nonumber
\end{align}
where $\omega_i^{\qty(p)}=\pm\sqrt{\qty(k_i^{\qty(p)})^2+m^2}$ are the vacuum energy eigenvalues.

We can construct the bases in 
a cylinder of radius $R$. 
This radius is ideally infinite, however, for numerical computation, it must take a finite value. %We set $R=25.0$ in the present paper. 
Imposing boundary conditions
\begin{align}
    J_{l^{\qty(p)}}\qty(k_i^{\qty(p)}R)&=0
\end{align}
for $l^{\qty(p)}=\kappa-\frac{3}{2},\ \kappa-\frac{1}{2},\ \text{or}\ \kappa+\frac{1}{2}$, we obtain some sets of discretized wave numbers $k_i^{\qty(p)},p=\qty(u,d,s)$. 
The orthogonality conditions are given by 
\begin{align}
    &\int_0^{R}\dd{\rho}\rho J_{l^{\qty(p)}}\qty(k_i^{\qty(p)}\rho)J_{l^{\qty(p)}}\qty(k_j^{\qty(p)}\rho)\nonumber\\
    &=\int_0^{R}\dd{\rho}\rho J_{{l^{\qty(p)}}\pm1}\qty(k_i^{\qty(p)}\rho)J_{{l^{\qty(p)}}\pm1}\qty(k_j^{\qty(p)}\rho)\nonumber\\
    &=\delta_{ij}\frac{R^2}{2}\qty[J_{{l^{\qty(p)}}\pm1}\qty(k_i^{\qty(p)}R)]^2.
\end{align}
Then, the normalization constants $\mathcal{N}_i^{(p)}$ are determined as
\begin{align}
    \mathcal{N}_i^{(p)}&=\qty[\frac{2\pi R^2\abs{\omega_i^{\qty(p)}}}{\abs{\omega_i^{\qty(p)}}+m}\qty(J_{l^{\qty(p)}}\qty(k_i^{\qty(p)}R))^2]^{-\frac{1}{2}}.
\end{align}
In this setup, the plane waves form orthonormal bases.

By solving the secular equation~\eqref{Eq:Secular} numerically for each parameter $\qty(a,b)$, we can obtain some energy spectra $\qty{\varepsilon}$ and corresponding wave functions $\qty{\psi}$ for each $\kappa$. 
The eigenproblem~\eqref{Eq:Secular} is solved by a standard matrix diagonalization solver 'dsyev' in the LAPACK, Math Kernel Library.
Note that to study the localization of wave functions to the background soliton in real space, we calculate the fermion density which is defined by 
\begin{equation}
	\chi_{\kappa}\qty(\rho)\equiv\frac{1}{2\pi}\int_0^{2\pi}\dd{\phi}\psi_{\kappa}^\dagger\psi_{\kappa}.
\end{equation}
In Sec.~\ref{Sec:Result_Density}, we shall see property of the localization of $\chi$ is strongly linked to the spectral flow.
For our numerical computations, we set $m=2.0, R=25.0$ and the number of discrete wavenumbers is chosen as $128$, which
gives a sufficient convergence.

%%%%%%%%%%%%%%%%%%%%%%%%%%%%Fig.5%%%%%%%%%%%%%%%%%%%%%%%%%%%%
\begin{figure*}[h]
	\centering
		\includegraphics[width=1\linewidth]{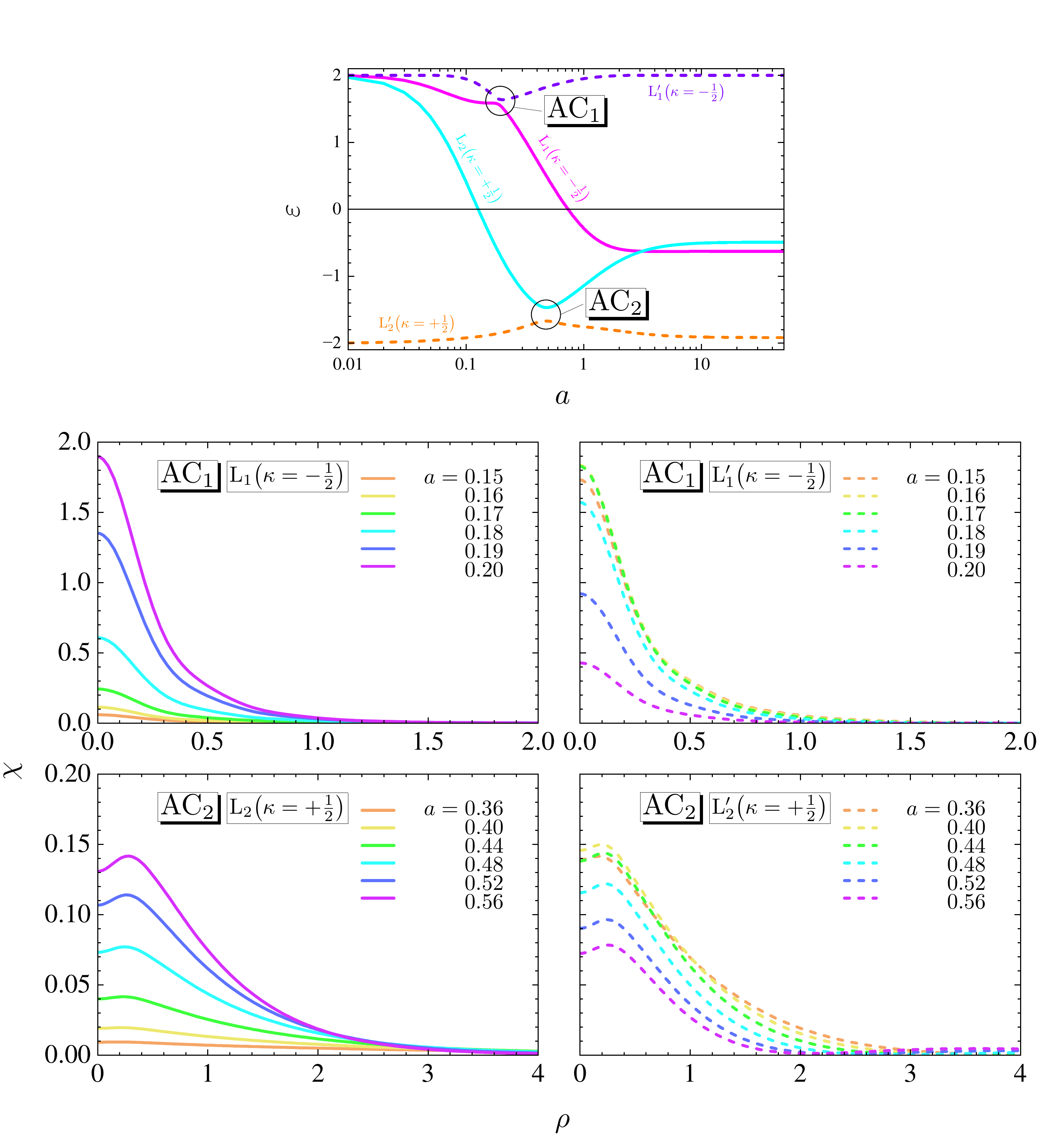}
		\caption{	
		The spectral flow of the anti-instanton by changing $a$ and corresponding fermion densities concerning the 
		ACs, the avoided crossings, which are the specific level repulsions with the exchange of the localization property. 		
		In the figure of the spectral flow, only the levels crossing the zero-energy line and the partner of the avoided crossing are shown. 
		The figures of fermion densities shows the fermion densities with their quantum number related to the AC. 
		%The correspondence between the color of the levels and the densities is as follows. AC$_1$: (left,right)=(magenta,violet), AC$_2$: (left,right)=(cyan,orange)
  		}
		\label{Fig:ZAaflow_AC}
\end{figure*}
%%%%%%%%%%%%%%%%%%%%%%%%%%%%Fig.5%%%%%%%%%%%%%%%%%%%%%%%%%%%%

\section{Results: the spectral flow}
\label{Sec:Result_flow}

Many previous studies on spectral flow analysis have shown the level crossings in which the background 
fields do not represent the solution of the model~\cite{Kahana:1984be,Kahana:1984dx,Witten:1982fp,Amari:2019tgs}.
One of the characteristic features of this study is that the background field is always the solution of the equation of motion,   
because of the spectral flow with the moduli parameters $\qty(a,b)$. 
Since changing $\qty(a,b)$ causes expansions or deformations of the solutions~(see Fig.~\ref{Fig:Qb1} and Fig.~\ref{Fig:Qa1}),
it is possible to derive the spectral flow as a function of these moduli parameters. 
As the spectral flow analysis with the instanton solutions was previously discussed in Ref.\cite{Amari:2023gjq}, 
here we concentrate on the results of the anti-instanton and the mixture solution.

Fig.\ref{Fig:ZAflow} plots the behavior of the energy spectra for the anti-instantons
with the parameters $(a,b)$ change. First, we begin by analyzing the result of varying $a$ 
shown in Fig.\ref{Fig:ZAflow}(a).
The topological charge $Q_A=0$ at $a=0$ and the background field is a vacuum solution; 
The number of positive and negative energy levels is strictly equal, 
as indicating the fermion number $N_f =0$. 
The topological charge switches on $Q_A=-2$ for $a\neq 0$. 
As gradually increasing the parameter $a$, the two energy levels (colored in magenta and cyan) 
move from the positive continuum to the negative. 
As a result, the number of positive levels decreases and negative rises by $2$ respectively, 
which corresponds to $N_\textrm{as}=4$, and then $N_f =2$. 
Note that $N_f$ and $Q_A$ are not necessarily correlated with each other. 
We see the local behavior of the energy levels of the regions I -- V in Fig.\ref{Fig:ZAflow}(a). 
From region I to II, the positive level decreases and the negative increases by 1 which corresponds to the change of $N_f:0\to 1$. 
From region II to III, the additional level comes from positive to negative, that is $N_f$ is now $2$. 
As a result, the number of zero modes ($=$ the number of crossing zero energy lines) between regions I and III equals to $2$. 
For larger $a$, since the topological charge $Q_A$ continues to be $-2$ as shown in Table.~\ref{Table:Qs}, 
%approaches the limit which is shown in Table.~\ref{Table:Limits}, and, because of $Q_A=-2$ at $a=\infty$, 
it is expected that no further net crossing can appear. 
Actually, in Fig.~\ref{Fig:ZAflow}(a), one can confirm that the sign flip of the energy levels does not occur at large $a$.\par

It is worth discussing the behavior of the soliton at the vicinity of $a\sim 0$. 
As it was already utilized in previous analysis~\cite{PhysRevLett.54.631,Kahana:1984be,Kahana:1984dx}, 
the solitons with tiny size do not induce the fermion number at all. 
The situation is comparable to the bound state problem in the quantum mechanical potential 
well~\cite{landau1981quantum}, when a bound state is absent below a certain threshold 
for the potential's size or depth. 
Therefore, the solution with sufficiently small $a$ has the same effect as the vacuum configuration. 

In the case with the parameter $b$~(Fig.\ref{Fig:ZAflow}(b), more complex behavior emerges.  
Given that the series passes over $Q_A=0\to -2\to -1$, there needs to be one more crossing zero energy. 
The magenta level descends and after that, it returns to the positive continuum. 
The transition of the topological charge appropriately results in this swing- back behavior, 
and the spectral flow appropriately connects the fermion number $N_f:~0\to 2\to 1$. 

Next, we discuss the spectral flow for the mixture, sphaleron solution $Z_M$. 
This solution has zero net topological charge $Q_M=0$ since it appears to composite two instantons and two anti-instantons. 
As we showed in Table \ref{Table:Qs}, it behaves as (anti-) instantons at some parameter limits. 
Therefore, the spectral flow connects three different states: the anti-instanton, the mixture, and the instanton.
As you can see in Fig.\ref{Fig:ZMflow}, four zero modes manifest in the process of the spectral flows. 
We begin with the change of $a$ in Fig.\ref{Fig:ZMflow}(a). 
With $a=0$ the topological charge $Q_M$ is $-2$, meaning that there are two more negative energy levels than there are 
in the vacuum state. 
As increasing $a$, the magenta levels and the cyan cross the zero-energy line from negative to positive, 
corresponding to the change of $N_f: 2\to0$. 
In region III, the number of negative levels and the positive levels are equal, i.e., $N_f=0$, 
and the energy gap between them significantly decreases,  
there is a nontrivial modulation to the vacuum. 
Further increasing $a$, additional two levels, the red and the blue move from negative to positive, 
which corresponds to the change of $N_f: 0\to -2$. 
In region V, the charge finally arrives $Q_M=2$ at $a=\infty$. 
Since the topological charge changes from $-2$ to $2$, the four zero modes manifest in the process of the spectral flow. 

As one can see in Fig.\ref{Fig:ZMflow}(b), 
again the behavior for the change of the parameter $b$ is a little more complex. 
The red level crosses the zero-energy line once and then crosses it again and goes back to the negative as growing $b$.  
In region I, the number of negative levels has increased by one. 
As increasing $b$, the red and cyan levels move from negative to positive, with the change of $N_f$: $-1\to0\to 1$. 
A larger $b$, the red level crosses the zero-energy line again, and the blue level 
moves from negative to positive, corresponding to $N_f$: $1\to0\to 1$. 
In this process, the energy spectra cross the zero-energy line a total of four times, indicating the existence of four zero modes. 
However, the red level undergoes an even number of crossings, resulting in net two crossings over the entire spectral flow.\par

In conclusion, we obtain a firm result regarding the complete correlation among the number of zero modes in all ranges of the parameters 
$\qty(a,b)$, the number of crossings of the zero energy line, and the change of $Q$ or $N_f$. 
The net number of crossings equals the difference in topological charges at the parameter limits. 
On the other hand, the number of zero modes is not necessarily equivalent to them 
as seen in Fig.\ref{Fig:ZAflow}(b) and Fig.\ref{Fig:ZMflow}(b). 
To fix the exact number of the zero modes, we just count the number of times the fermion number changes. 
The spectral flow analysis is the best way to do this. 

\section{Results: The localizing Fermions}
\label{Sec:Result_Density}

The localizing Dirac fermions coupled with the texture-type topological soliton (i.e.,the skyrmion) 
were first addressed in Refs.\cite{Kahana:1984be,Kahana:1984dx}. 
The zero modes of fermions are localized on a linear profile function: 
$f(r)\sim \pi-\pi r/\xi,~0\leq r\leq \xi$ of sufficiently large size $\xi$.  
We stress once more that they do not represent the solutions on the genuine background profile. 
In this section, we present the true localizing solutions on the genuine solitons.
\par

First, we consider the case of the anti-instanton.
At the vacuum, $Q_A=0$ occurs with $a=0$ or $b=0$ (see Table.~\ref{Table:Qs}), 
and the fermions are not localized. %do not localize on. 
As increasing the parameters $a, b$, the localization emerges.  
It is an intricate process that is often triggered by the energy level junction, 
which is also referred to as an energy level repulsion or an avoided crossing (AC). 
Fig.~\ref{Fig:ZAaflow_AC} shows the spectral flow and the fermion densities of varying $a$ 
for the anti-instanton. 
The levels concerning the avoided crossing are L$_i$, L$_i',i=1,2$, and they are situated 
at AC$_1$ and AC$_2$. 
On the junctions, the localization property is exchanged with 
each other. For example, at around the AC$_1$, below $a\sim 0,4$ the density of L$_1'$ is localized, while the L$_1$ is not.
The L$_1'$ density starts to delocalize and the L$_1$ density localizes at $a\geq 0.4$.  
The same has happened at the AC$_2$. 

\begin{figure*}[t]
	\centering
		\includegraphics[width=1\linewidth]{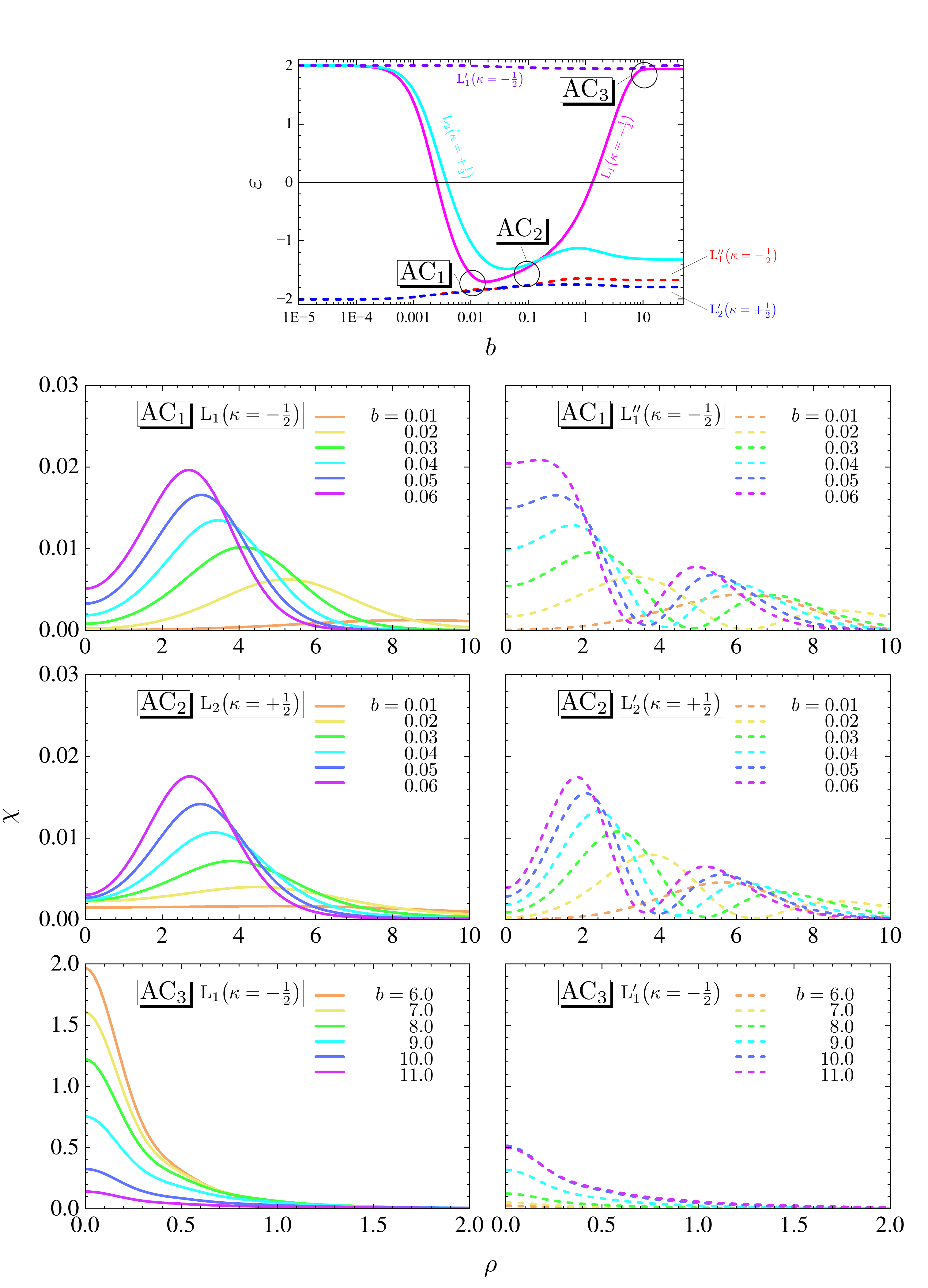}
		\caption{The spectral flow of the anti-instanton by changing $b$ and corresponding fermion densities around the ACs, 
		the avoided crossings. 
		In the figure of the spectral flow, only the levels crossing the zero-energy line and the partner of the avoided crossing are shown. 
		The figures of fermion densities shows the fermion densities with their quantum number related to the AC. 
		It should be noted that L$_1''$ and L$_2'$ oscillate because of their respective delocalizations.		
		%The correspondence between the color of the levels and the densities is as follows. AC$_1$: (left,right)=(magenta,red), 
		AC$_2$: (left,right)=(cyan,blue), AC$_3$: (left,right)=(magenta,violet)
  	}
		\label{Fig:ZAbflow_AC}
\end{figure*}

In Fig.~\ref{Fig:ZAbflow_AC}, the spectral flow for the change of $b$ is plotted and, 
one can see some ACs.  
The L$_1$,L$_2$ are the delocalized mode at $b\lesssim 10^{-4}$. 
They transition into localized modes via the AC$_1$ at $b\sim 0.01$. 
It is a almost similar behavior as the previous one, but now the level L$_1$ again encounters the next ACs,  
interaction between levels L$_1$ and L$_1'$ occurs at $b\sim10$.
Now the localization of the L$_1$ transitions to the counterpart L$_1'$.  
The L$_1'$ is to be a continuous mode in the limit of $b\to\infty$, by the additional AC with the other continuous levels. 
Since the L$_2$ remains a discrete level for the whole $b$.
Consequently, only the density of L$_2$ remains as the localizing mode for $b$ large, that is, $b\sim100$.
\par

\begin{figure*}[t]
	\centering
		\includegraphics[width=1\linewidth]{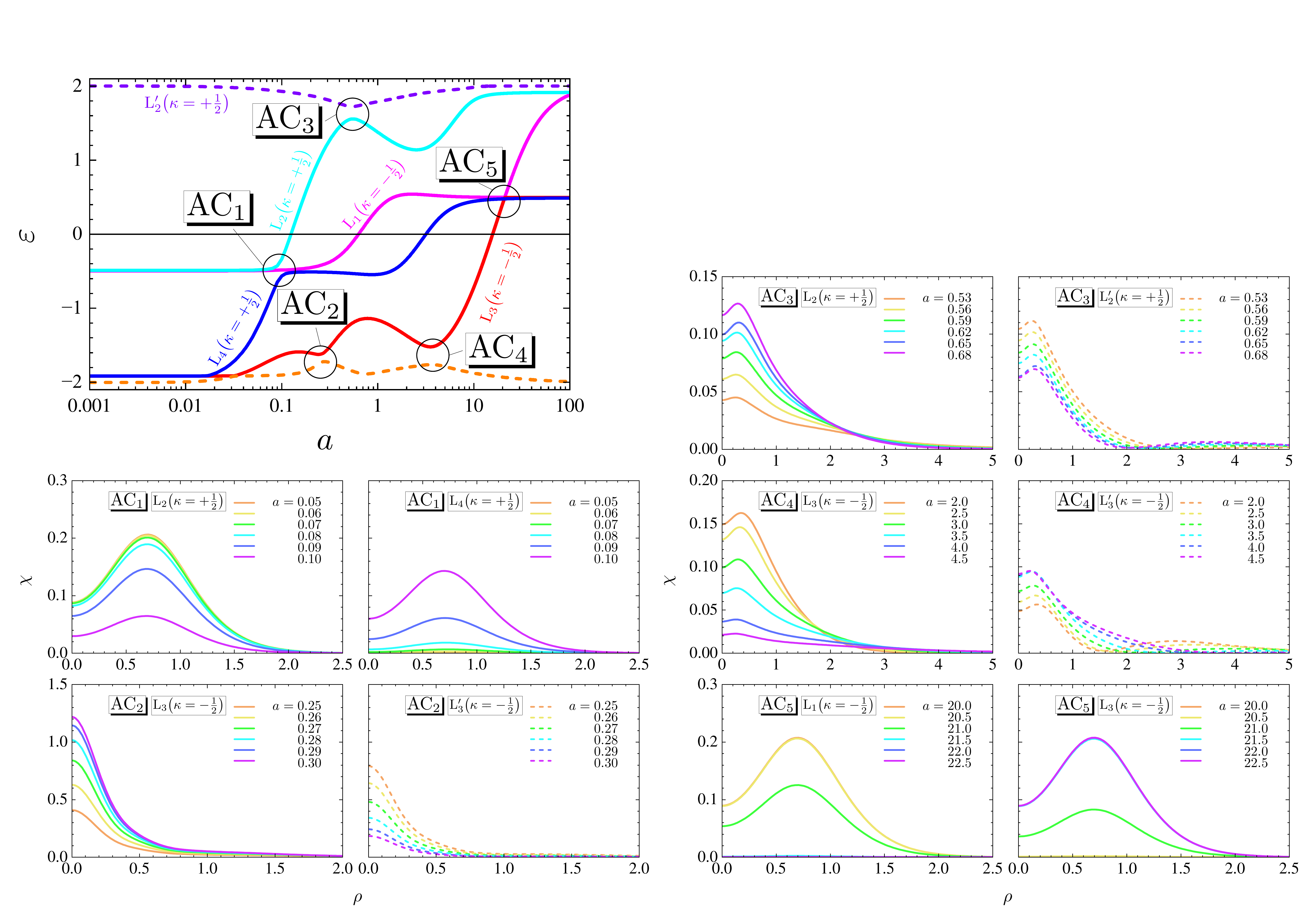}
		\caption{The spectral flow of the mixture by changing $a$ and corresponding fermion densities around the ACs, 
		the avoided crossings.  
		In the figure of the spectral flow, only the levels crossing the zero-energy line and the partner of the avoided crossing are shown. 
		The figures of fermion densities shows the fermion densities with their quantum number related to the AC. 
		%The correspondence between the color of the levels and the densities is as follows. AC$_1$: (left,right)=(cyan,blue), AC$_2$: (left,right)=(red,orange)
  }
		\label{Fig:ZMaflow_AC}
\end{figure*}

\begin{figure*}[h]
	\centering
		\includegraphics[width=1.0\linewidth]{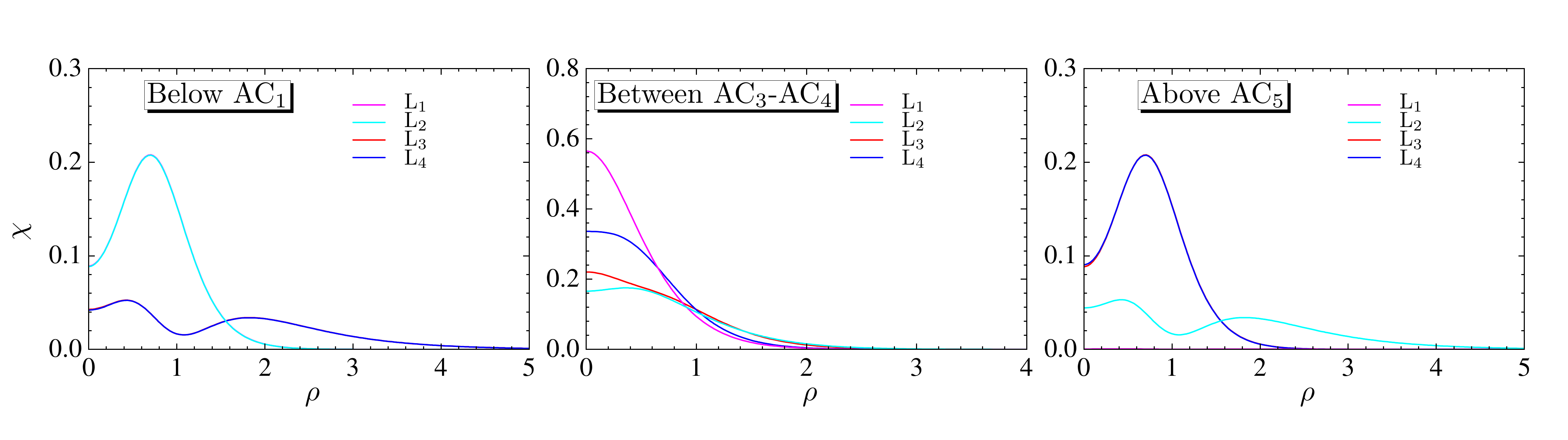}
		\caption{The typical behaviors of fermion densities in terms of L$_1$, L$_2$, L$_3$ and L$_4$, 
		concerning the five types of the avoided crossings. 
		For small $a$, below the AC$_1$, the densities of L$_1$ and L$_2$ are localized. 
		In intermediate regions, between the AC$_3$ and AC$_4$, all of the densities are localized. 
		For large $a$, above the AC$_5$, the densities of L$_3$ and L$_4$ are localized.}
		\label{Fig:ZMaflow_bba}
\end{figure*}

Our new insights into the fermion density localization are effective for understanding the case of the mixture solution. 
Fig.~\ref{Fig:ZMaflow_AC} shows the spectral flow of the mixture and the corresponding fermion density as $a$ is varied. 
The four levels move from negative to positive as $a$ grows, and the five ACs switch their localization and delocalization. 
This results in the localization of L$_1$ and L$_2$ densities in the region where $a$ is small, and of L$_3$ and L$_4$ densities in the region where $a$ is large. 
Recalling that the mixture behaves as the anti-instanton where $a$ is sufficiently small and as the instanton where $a$ is sufficiently large, one can see that the L$_1$ and L$_2$ densities are modes localized in the anti-instanton part of the mixture, and the L$_3$ and L$_4$ densities in the instanton part. 
Here, the behavior of the fermion density around $a=1$ gives us a deep understanding of the mixture solution. 
In the region, the four fermion densities are localized as shown in Fig.~\ref{Fig:ZMaflow_bba}. 
Consequently, the modes localized in the anti-instanton component and the instanton part coexist in this region. 
This means that the mixture with $Q_M=0$ topological charge is not just a topologically trivial field, but has the function of trapping fermions and closing the energy gap as discussed in the previous section. 
Moreover, we can interpret fermions can detect the instantons and anti-instantons inside the mixture.

\begin{figure*}[t]
	\centering
		\includegraphics[width=1.0\linewidth]{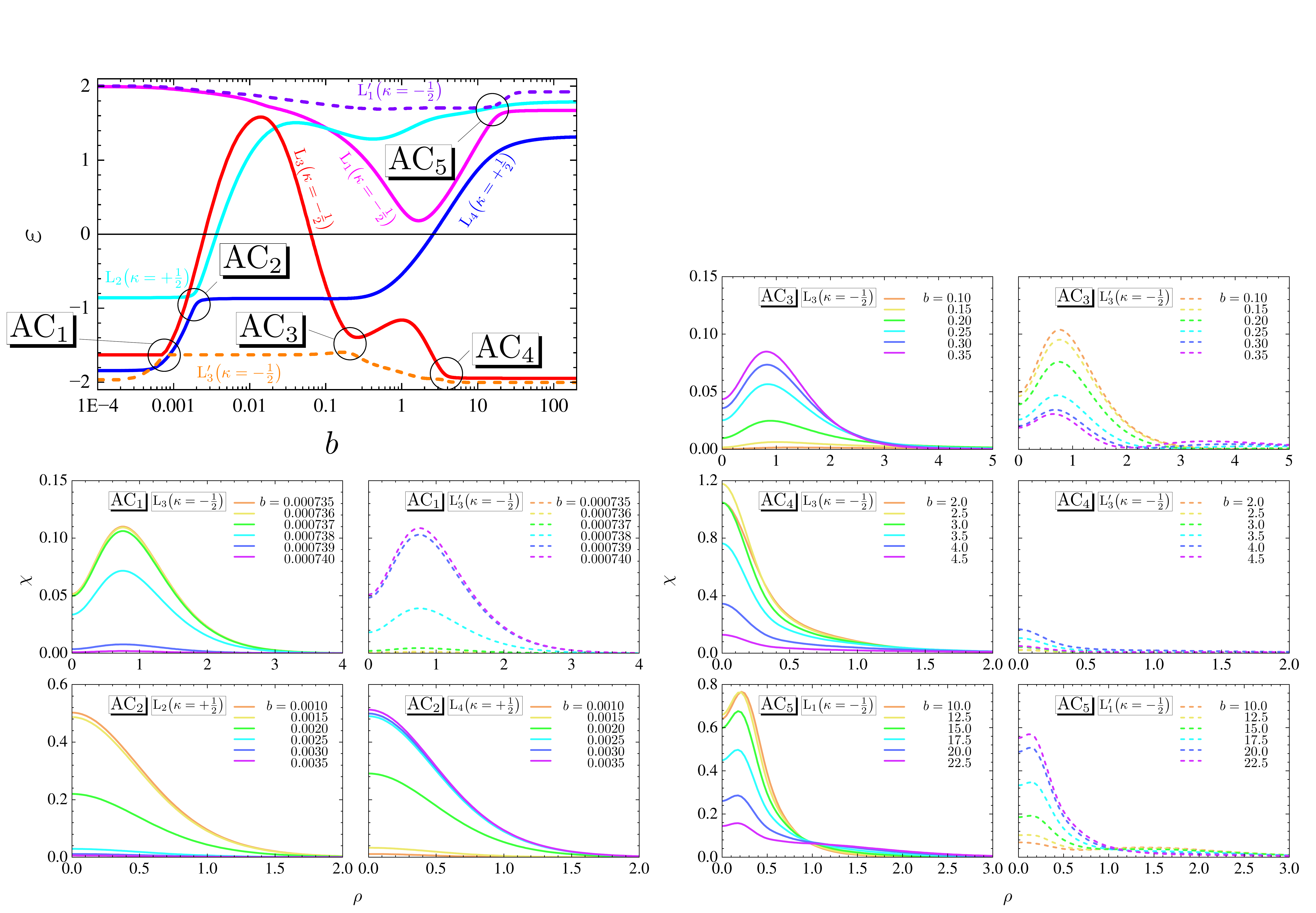}
		\caption{The spectral flow of the mixture by changing $b$ and corresponding fermion densities around the ACs, 
		the avoided crossings.  
		In the figure of the spectral flow, only the levels crossing the zero-energy line and the partner of the avoided crossing are shown. 
		The figures of fermion densities shows the fermion densities with their quantum number related to the AC. 
		%The correspondence between the color of the levels and the densities is as follows. AC$_1$: (left,right)=(red,orange), AC$_2$: (left,right)=(cyan,blue)
  }
		\label{Fig:ZMbflow_AC}
\end{figure*}

\begin{figure*}[h]
	\centering
		\includegraphics[width=1.0\linewidth]{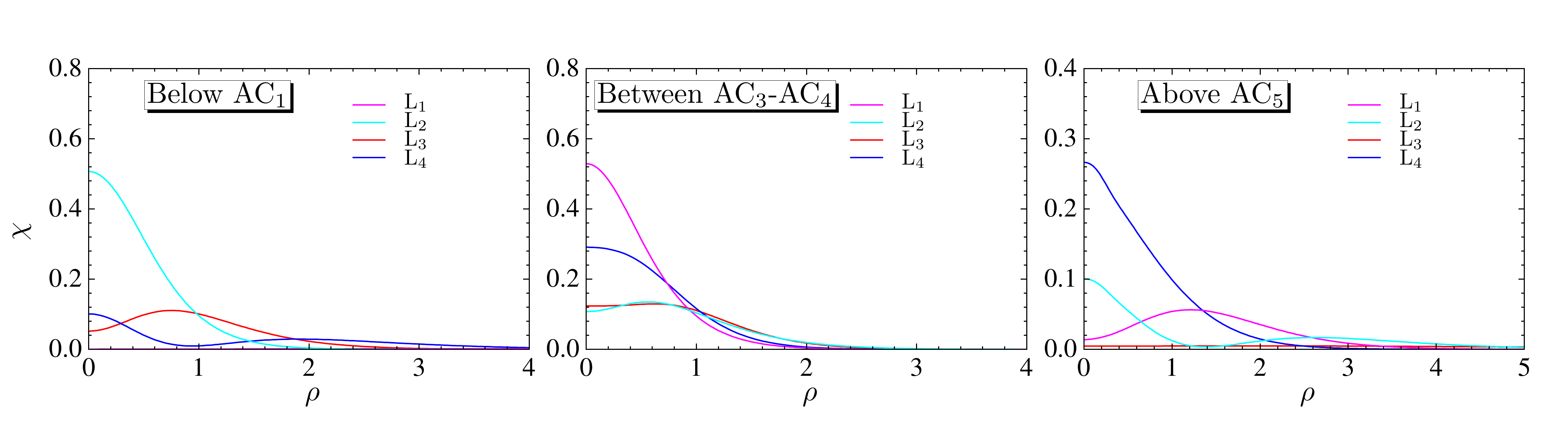}
		\caption{The typical behaviors of fermion densities in terms of L$_1$, L$_2$, L$_3$ and L$_4$ 
		concerning the five types of the avoided crossings. 
		For small $b$, below the AC$_1$, the densities of L$_2$ and L$_3$ are localized. In intermediate regions, 
		between the AC$_3$ and AC$_4$, all of the densities are localized. For large $b$, above the AC$_5$, 
		the densities of L$_1$ and L$_4$ are localized.}
		\label{Fig:ZMbflow_bba}
\end{figure*}

The (de)localization of fermion densities occur by the same mechanism in the spectral flow when $b$ is varied, 
as shown in Fig.~\ref{Fig:ZMbflow_AC} and Fig.~\ref{Fig:ZMbflow_bba}. 
Since the %red colored 
level L$_3$ crosses the zero-energy line an even number of times, the switching between localized and 
delocalized state occurs more than once (totally three times), as explained in the anti-instanton case.

%%%%%%%%%%%%%%%%%%%%%%%%%%%%%%%%%%%%%%%%%%%%%%%%%%%%%%%%%%%%%%%%%%%%%%
%CONCLUSIONS

\section{Conclusions}
\label{Sec:Conclusions}

In this paper, we developed a novel approach to the realization of spectral flows pertaining to solitons' parameters in the $\CP^2$ NLSM. 
We elucidated the connection between fermion density localization and the emergence of the fermion zero mode. 
Since the (anti-)instantons and the mixtures are the exact solutions with the two moduli parameters, our spectral flow 
analysis meaningful for the whole parameter space. 
At the spectral flow, some energy levels cross zero and the net number of the crossings coincides with 
the difference of topological charge of both ends of the flow.
In the previous studies concerning the spectral flow, they claimed that 
the number of crossings should be equivalent to the number of zero modes. 
In this paper, we examined the flows between the solutions with the several charges, so
the number of zero modes reflect how many times the topological charge changes its value.

The solutions consist of the anti-instanton ($Q_M<0$), the mixture ($Q_M=0$) and the instanton ($Q_M>0$). 
The spectral flow analysis for the mixture is one of the most intriguing results. 
The energy gap closes in the $Q_M=0$ region, indicating that the mixture realizes an 
instanton + anti-instanton composite with an interior structure rather than 
just being a topologically trivial one.

For the localization property of the fermions, generally, a large soliton solution induces the 
localizing fermion mode.  
It is also essential that the transition between localizing and delocalizing fermion density is observed.  
The avoided crossings between relevant energy levels play a role in such transition of the fermion density. 
In other words, it is concluded that the localization of the fermion density depends not 
only on the soliton size but also on the presence of AC in the spectral flow.
The localization also occurs in the spectral flow of the mixture solution. 
Of particular interest is the fact that the four fermion densities are localized in the region $Q_M=0$, 
containing both the instanton-localized and the anti-instanton-localized densities. 
This result implies that the fermions can successfully detect the (anti-) instanton inside the mixture. 

Our spectral flow analysis, utilizing parameters of the solutions, has provided two crucial outcomes: 
\begin{itemize}
\item[1.] The spectral asymmetry,
\item[2.] The (de)localization of density of the fermions. 
\end{itemize}
Both are deeply intertwined with the topology through the index theorem. 
The former emerges as a manifestation of the global change of the topology, 
represented by the difference in topological charges at the outside of the parameter space. (see Sec.V). 
On another hand, the latter has a more geometrical origin; it often arises through the avoided crossings in the spectral flow process. 
It is worth repeating that we were able to discuss the second property since the spectral flow was treated exactly. 
Since topology signifies a solitonic symmetry, the avoided crossings indicate signs of subtle modulation of the symmetry. 
Therefore, it is envisaged that the avoided crossings arising in the spectrum flow will 
serve as a harbor for new physics in the fermion-soliton system.   
As case studies, the avoided crossings could be explored in systems such as the flag manifold sigma model
~\cite{AMARI2018294,PhysRevD.97.065012,Bykov:2015pka,Kobayashi:2021qfj}, an extension of the $\CP^{N-1}$ NLSM to large homotopy group, 
or composite soliton systems such as domain-wall skyrmion chain~\cite{PhysRevD.89.085022}.
We will report on these issues in our subsequent papers.

%%%%%%%%%%%%%%%%%%%%%%%%%%%%%%%%%%%%%%%%%%%%%%%%%%%%%%%%%%%%%%%%%%%%%%
%ACKNOWLEDGEMENT

\section*{Acknowledgement}
The authors would like to thank P.Klimas, Y.Shnir, F.Blaschke, F.Hanada, A.Nakamula and K.Toda for their many usefull comments.
Y.A. was supported in part by JSPS KAKENHI Grant Number JP23KJ1881.
N.S. was supported in part by JSPS KAKENHI Grant Number JP20K03278, and also JP23K02794.

%%%%%%%%%%%%%%%%%%%%%%%%%%%%%%%%%%%%%%%%%%%%%%%%%%%%%%%%%%%%%%%%%%%%%%

%%%%%%%%%%%%%%%%%%%%%%%%%%%%Fig.6%%%%%%%%%%%%%%%%%%%%%%%%%%%%
%%%%%%%%%%%%%%%%%%%%%%%%%%%%Fig.6%%%%%%%%%%%%%%%%%%%%%%%%%%%%

%%%%%%%%%%%%%%%%%%%%%%%%%%%%Fig.7%%%%%%%%%%%%%%%%%%%%%%%%%%%%
%%%%%%%%%%%%%%%%%%%%%%%%%%%%Fig.7%%%%%%%%%%%%%%%%%%%%%%%%%%%%

%%%%%%%%%%%%%%%%%%%%%%%%%%%%Fig.8%%%%%%%%%%%%%%%%%%%%%%%%%%%%
%%%%%%%%%%%%%%%%%%%%%%%%%%%%Fig.8%%%%%%%%%%%%%%%%%%%%%%%%%%%%

%%%%%%%%%%%%%%%%%%%%%%%%%%%%Fig.9%%%%%%%%%%%%%%%%%%%%%%%%%%%%
%%%%%%%%%%%%%%%%%%%%%%%%%%%%Fig.9%%%%%%%%%%%%%%%%%%%%%%%%%%%%

%%%%%%%%%%%%%%%%%%%%%%%%%%%%Fig.10%%%%%%%%%%%%%%%%%%%%%%%%%%%%
%%%%%%%%%%%%%%%%%%%%%%%%%%%%Fig.10%%%%%%%%%%%%%%%%%%%%%%%%%%%%

%APPENDIX

\appendix

\section{Derivation of the mixed charge or energy density}
\label{AppendixA}
In this appendix, we derive the topological charge or energy density of a $\CP^2$ mixture in terms of the ( anti-) instanton charge density $\mathcal{Q}_I$ and $\mathcal{Q}_A$ or energy density $\mathcal{E}_I$ and $\mathcal{E}_A$. We use the following properties without proof:
\begin{align}
	P_+^k f\bot P_+^{k'} f,\ k\neq k',\\
	P_+^0f=f,\ P_+^3f=0.
\end{align}
The first one can be verified recursively, and one can directly see that the second one is automatically satisfied for an arbitrary $f\in\mathbb{C}^3\backslash\qty{0}$.

Let us define a B\"acklund transformation:
\begin{align}
	P_\pm f&=\partial_\pm f-\frac{f^\dagger\cdot\partial_\pm f}{\abs{f}^2}f,\\
	P_\pm^kf&=\partial_\pm P_\pm^{k-1}f-\frac{\qty(P_\pm^{k-1}f)^\dagger\cdot\partial_\pm P_\pm^{k-1}f}{\abs{P_\pm^{k-1}f}^2}P_\pm^{k-1}f.
\end{align}
This definition leads a relation, $D_\pm\dfrac{g}{\abs{g}}=\dfrac{1}{\abs{g}}P_\pm g$
%\begin{align}
%	D_\pm\frac{g}{\abs{g}}
%	&=\frac{1}{\abs{g}}P_\pm g
%\end{align}
for an arbitrary $g\in\mathbb{C}^3\backslash\qty{0}$. Therefore, the topological charge densities have the form below:
\begin{align}
	\mathcal{Q}_I\qty(x)&=\mathcal{Q}_{\qty(0)}\qty(x)=2\qty[\abs{D_+Z_{\qty(0)}}^2-\abs{D_-Z_{\qty(0)}}^2]\nonumber\\
	&=\frac{2}{\abs{f}^2}\qty(\abs{P_+f}^2-\abs{P_-f}^2)
	=2\frac{\abs{P_+f}^2}{\abs{f}^2},\\
	\mathcal{Q}_M\qty(x)&=\mathcal{Q}_{\qty(1)}\qty(x)=2\qty[\abs{D_+Z_{\qty(1)}}^2-\abs{D_-Z_{\qty(1)}}^2]\nonumber\\
	&=\frac{2}{\abs{P_+f}^2}\qty(\abs{P_+^2f}^2-\abs{P_-P_+f}^2),\\
	\mathcal{Q}_A\qty(x)&=\mathcal{Q}_{\qty(2)}\qty(x)=2\qty[\abs{D_+Z_{\qty(2)}}^2-\abs{D_-Z_{\qty(2)}}^2]\nonumber\\
	&=\frac{2}{\abs{P_+^2f}^2}\qty(\abs{P_+^3f}-\abs{P_-P_+^2f}^2)\nonumber\\
	&=-2\frac{\abs{P_-P_+^2f}^2}{\abs{P_+^2f}^2}.
\end{align}
In the same way, the energy densities have the form below:
\begin{align}
	\mathcal{E}_I\qty(x)&=\mathcal{E}_{\qty(0)}\qty(x)=2\frac{\abs{P_+f}^2}{\abs{f}^2},\\
	\mathcal{E}_M\qty(x)&=\mathcal{E}_{\qty(1)}\qty(x)=\frac{2}{\abs{P_+f}^2}\qty(\abs{P_+^2f}^2+\abs{P_-P_+f}^2),\\
	\mathcal{E}_A\qty(x)&=\mathcal{E}_{\qty(2)}\qty(x)=2\frac{\abs{P_-P_+^2f}^2}{\abs{P_+^2f}^2}.
\end{align}
Our goal is to derive two relations:
\begin{align}
	\frac{\abs{P_-P_+f}^2}{\abs{P_+f}^2}&=\frac{\abs{P_+f}^2}{\abs{f}^2},\\
	\frac{\abs{P_-P_+^2f}^2}{\abs{P_+^2f}^2}&=\frac{\abs{P_+^2f}^2}{\abs{P_+f}^2}.
\end{align}
These relations lead $\mathcal{Q}_M=\abs{\mathcal{Q}_A}-\mathcal{Q}_I$ and $\mathcal{E}_M=\abs{\mathcal{E}_A}+\mathcal{E}_I$.
%\begin{align}
%	\mathcal{Q}_M&=\abs{\mathcal{Q}_A}-\mathcal{Q}_I,\\
%	\mathcal{E}_M&=\abs{\mathcal{E}_A}+\mathcal{E}_I.
%\end{align}

\subsection{Instanton part}
In this subsection, we show 
\begin{align}
	\frac{\abs{P_-P_+f}^2}{\abs{P_+f}^2}=\frac{\abs{P_+f}^2}{\abs{f}^2}.
\end{align}
In order to see this, we calculate
\begin{align}
	P_-P_+f=\partial_-P_+f-\frac{\qty(P_+f)^\dagger\cdot\partial_-P_+f}{\abs{P_+f}^2}P_+f.\label{PmPpf}
\end{align}
The second term of Eq.~\eqref{PmPpf} vanishes because of $P_+f\bot f$,
\begin{align}
	&\qty(P_+f)^\dagger\cdot\partial_-P_+f\nonumber\\
	&=\frac{1}{\abs{f}^2}\qty(-\abs{\partial_+f}^2+\frac{\abs{f^\dagger\cdot\partial_+f}^2}{\abs{f}^2})\qty(P_+f)^\dagger\cdot f\nonumber\\
	&=0.
\end{align}
Thus we obtain 
\begin{align}
	P_-P_+f=\partial_-P_+f=\frac{1}{\abs{f}^2}\qty(-\abs{\partial_+f}^2+\frac{\abs{f^\dagger\cdot\partial_+f}^2}{\abs{f}^2})f.\label{PmPpfresult}
\end{align}
Here we consider $\abs{P_+f}^2
=\abs{\partial_+f}^2-\dfrac{\abs{f^\dagger\cdot\partial_+f}^2}{\abs{f}^2}$. %the norm of $P_+f$.
%\begin{align}
%	\abs{P_+f}^2
%	&=\abs{\partial_+f}^2-\frac{\abs{f^\dagger\cdot\partial_+f}^2}{\abs{f}^2}.
%\end{align}
Then, $P_-P_+f$ has the form
\begin{align}
	P_-P_+f
	&=-\frac{\abs{P_+f}^2}{\abs{f}^2}f,
\end{align}
and we have%its norm is 
\begin{align}
	\abs{P_-P_+f}^2
	&=\frac{\abs{P_+f}^4}{\abs{f}^2}.
\end{align}
Therefore, we obtain
\begin{align}
	\frac{\abs{P_-P_+f}^2}{\abs{P_+f}^2}=\frac{\abs{P_+f}^2}{\abs{f}^2}.\label{result1}
\end{align}

\subsection{Anti-instanton part}
In this subsection, we show 
\begin{align}
	\frac{\abs{P_-P_+^2f}^2}{\abs{P_+^2f}^2}=\frac{\abs{P_+^2f}^2}{\abs{P_+f}^2}.
\end{align}
The method is the same as for the instanton part. We calculate 
\begin{align}
	P_-P_+^2f=\partial_-P_+^2f-\frac{\qty(P_+^2f)^\dagger\cdot\partial_-P_+^2f}{\abs{P_+^2f}}P_+^2f.\label{PmPpPpf}
\end{align}
The second term of Eq.~\eqref{PmPpPpf} vanishes because of $P_+^2f\bot P_+f,\ P_+^2f\bot f$. First, we show this fact. The main component can be reduced into
\begin{align}
	\partial_-P_+^2f
	&=\partial_+\partial_-P_+f-\frac{\abs{\partial_+P_+f}^2}{\abs{P_+f}^2}P_+f\nonumber\\
	&-\frac{\qty(P_+f)^\dagger\cdot\partial_+\partial_-P_+f}{\abs{P_+f}^2}P_+f\nonumber\\
	&+\frac{\qty(P_+f)^\dagger\cdot\partial_+P_+f}{\abs{P_+f}^4}\partial_-\abs{P_+f}^2P_+f\nonumber\\
	&-\frac{\qty(P_+f)^\dagger\cdot\partial_+P_+f}{\abs{P_+f}^2}\partial_-P_+f.\label{pmPpPpf}
\end{align}
Since we have
\begin{align}
	\partial_-P_+f=P_-P_+f=-\frac{\abs{P_+f}^2}{\abs{f}^2}f
\end{align}
from Eq.~\eqref{PmPpfresult}, the component of the third term of Eq.~\eqref{pmPpPpf} has the form
\begin{align}
	&\qty(P_+f)^\dagger\cdot\partial_+\partial_-P_+f\nonumber\\
	&=-\frac{\partial_+\abs{P_+f}^2}{\abs{f}^2}\qty(P_+f)^\dagger\cdot f\nonumber\\
	&+\frac{\abs{P_+f}^2}{\abs{f}^4}\partial_+\abs{f}^2\qty(P_+f)^\dagger\cdot f\nonumber\\
	&-\frac{\abs{P_+f}^2}{\abs{f}^2}\qty(P_+f)^\dagger\cdot\partial_+f\nonumber\\
	&=-\frac{\abs{P_+f}^4}{\abs{f}^2}.\label{3pmPpPpf}
\end{align}
And the component of the second term of Eq.~\eqref{pmPpPpf} is
\begin{align}
	\abs{\partial_+P_+f}^2
	&=\abs{P_+^2f}^2+\frac{\abs{\qty(P_+f)^\dagger\cdot\partial_+P_+f}^2}{\abs{P_+f}^2}.
\end{align}
As a result, Eq.~\eqref{pmPpPpf} can be written as follows:
\begin{align}
	\partial_-P_+^2f
	&=\partial_+\partial_-P_+f-\frac{\abs{P_+^2f}^2}{\abs{P_+f}^2}P_+f\nonumber\\
	&-\frac{\abs{\qty(P_+f)^\dagger\cdot\partial_+P_+f}^2}{\abs{P_+f}^4}P_+f+\frac{\abs{P_+f}^2}{\abs{f}^2}P_+f\nonumber\\
	&+\frac{\qty(P_+f)^\dagger\cdot\partial_+P_+f}{\abs{P_+f}^4}\partial_-\abs{P_+f}^2P_+f\nonumber\\
	&-\frac{\abs{\qty(P_+f)^\dagger\cdot\partial_+P_+f}^2}{\abs{P_+f}^2}\partial_-P_+f.\label{pmPpPpfresult}
\end{align}
Here using a relation
\begin{align}
	\partial_-\abs{P_+f}^2
	&=\qty(\partial_+P_+f)^\dagger\cdot P_+f,
\end{align}
Eq.~\eqref{pmPpPpfresult} has a more simple form,
\begin{align}
	\partial_-P_+^2f
	&=\partial_+\partial_-P_+f+\qty(-\frac{\abs{P_+^2f}^2}{\abs{P_+f}^2}+\frac{\abs{P_+f}^2}{\abs{f}^2})P_+f\nonumber\\
	&+\frac{\abs{\qty(P_+f)^\dagger\cdot\partial_+P_+f}^2}{\abs{f}^2}f.
\end{align}
Recalling that $\partial_+\partial_-P_+f$ does not contain $P_+^2f$ (see Eq.~\eqref{3pmPpPpf} and $\partial_+f=P_+f+\frac{f^\dagger\cdot\partial_+f}{\abs{f}^2}f$), the second term of Eq.~\eqref{PmPpPpf} vanishes, i.e.
\begin{align}
	\qty(P_+^2f)^\dagger\cdot\partial_-P_+^2f=0.
\end{align}
Therefore we obtain
\begin{align}
	&P_-P_+^2f\nonumber\\
	&=\partial_-P_+^2f\nonumber\\
	&=\partial_+\partial_-P_+f+\qty(-\frac{\abs{P_+^2f}^2}{\abs{P_+f}^2}+\frac{\abs{P_+f}^2}{\abs{f}^2})P_+f\nonumber\\
	&+\frac{\qty(P_+f)^\dagger\cdot\partial_+P_+f}{\abs{f}^2}f.\label{PmPpPpfresult}
\end{align}
From Eq.~\eqref{3pmPpPpf}, the first term of Eq.~\eqref{PmPpPpfresult} is 
\begin{align}
	&\partial_+\partial_-P_+f\nonumber\\
	&=-\frac{\partial_+\abs{P_+f}^2}{\abs{f}^2}f+\frac{\abs{P_+f}^2}{\abs{f}^4}\partial_+\abs{f}^2f-\frac{\abs{P_+f}^2}{\abs{f}^2}\partial_+f.
\end{align}
Considering 
\begin{align}
	\partial_+\abs{P_+f}^2
	&=\qty(P_+f)^\dagger\cdot\partial_+P_+f,
\end{align}
Eq.~\eqref{PmPpPpfresult} can be written as 
\begin{align}
	P_-P_+^2f
	&=-\frac{\abs{P_+^2f}^2}{\abs{P_+f}^2}P_+f,
\end{align}
and then we have%the norm is 
\begin{align}
	\abs{P_-P_+^2f}^2
	&=\frac{\abs{P_+^2f}^4}{\abs{P_+f}^2}.
\end{align}
Therefore, we obtain the relation
\begin{align}
	\frac{\abs{P_-P_+^2f}^2}{\abs{P_+^2f}^2}=\frac{\abs{P_+^2f}^2}{\abs{P_+f}^2}.\label{result2}
\end{align}
Then, from Eq.~\eqref{result1} and Eq.~\eqref{result2}, we find $\abs{D_+Z_{\qty(1)}}^2=\abs{D_-Z_{\qty(2)}}^2$ and $\abs{D_-Z_{\qty(1)}}^2=\abs{D_+Z_{\qty(0)}}^2$.

\bibliography{ref}

\end{document}